\documentclass[iop]{emulateapj}  

\usepackage{color}
\usepackage{hyperref}  
\usepackage{breakurl}  
\usepackage{lineno}
\usepackage{comment}

\newcommand\kms{\ifmmode{\rm km\thinspace s^{-1}}\else km\thinspace s$^{-1}$\fi}

\shortauthors{Torres et al.}
\shorttitle{Castor}

\begin{document}
\submitted{Accepted for publication in The Astrophysical Journal}

\title{The Orbits and Dynamical Masses of the Castor System}

\author{
Guillermo Torres\altaffilmark{1}, 
Gail H.\ Schaefer\altaffilmark{2}, 
John D.\ Monnier\altaffilmark{3}, 
Narsireddy Anugu\altaffilmark{2}, 
Claire L.\ Davies\altaffilmark{4}, 
Jacob Ennis\altaffilmark{3}, 
Christopher D.\ Farrington\altaffilmark{2}, 
Tyler Gardner\altaffilmark{3}, 
Robert Klement\altaffilmark{2}, 
Stefan Kraus\altaffilmark{4}, 
Aaron Labdon\altaffilmark{5}, 
Cyprien Lanthermann\altaffilmark{2}, 
Jean-Baptiste Le Bouquin\altaffilmark{6}, 
Benjamin R.\ Setterholm\altaffilmark{3}, and 
Theo ten Brummelaar\altaffilmark{2} 
}

\altaffiltext{1}{Center for Astrophysics $\vert$ Harvard \&
  Smithsonian, 60 Garden St., Cambridge, MA 02138, USA;
  gtorres@cfa.harvard.edu}

\altaffiltext{2}{The CHARA Array of Georgia State University, Mount Wilson Observatory, Mount Wilson, CA 91203, USA}

\altaffiltext{3}{Astronomy Department, University of Michigan, Ann Arbor, MI 48109, USA}

\altaffiltext{4}{Astrophysics Group, Department of Physics \& Astronomy, University of Exeter, Stocker Road, Exeter, EX4 4QL, UK}

\altaffiltext{5}{European Southern Observatory, Casilla 19001, Santiago 19, Chile}

\altaffiltext{6}{Universit\'e Grenoble Alpes, CNRS, IPAG, 38000, Grenoble, France}

\begin{abstract} 

Castor is a system of six stars in which the two brighter objects,
Castor~A and B, revolve around each other every $\sim$450 yr and are
both short-period spectroscopic binaries. They are attended by the
more distant Castor~C, which is also a binary.  Here we report
interferometric observations with the CHARA array that spatially resolve the
companions in Castor~A and B for the first time. We complement these
observations with new radial velocity measurements of A and B spanning
30 yr, with the Hipparcos intermediate data, and with existing
astrometric observations of the visual AB pair obtained over the past
three centuries. We perform a joint orbital solution to solve
simultaneously for the three-dimensional orbits of Castor~A and B as
well as the AB orbit. We find that they are far from being coplanar:
the orbit of A is nearly at right angles (92\arcdeg) relative to the wide orbit,
and that of B is inclined about 59\arcdeg\ compared to AB.
We determine the dynamical masses of the four stars in Castor~A and B
to a precision better than 1\%.  We also determine the radii of the
primary stars of both subsystems from their angular diameters measured
with CHARA, and use them together with stellar evolution models to
infer an age for the system of 290~Myr. The new knowledge of the
orbits enables us to measure the slow motion of Castor~C as well,
which may assist future studies of the dynamical evolution of this
remarkable sextuple system.

\end{abstract}

\section{Introduction}
\label{sec:introduction}

Castor ($\alpha$~Geminorum) is a bright and well-known nearby star system only 15~pc
away that has been followed as a visual binary for more than three
centuries. Its discovery as such is credited to the English astronomers James
Pound and James Bradley \citep[see][]{Herschel:1833}. They first observed it
in 1718 and 1719, respectively, although it appears that
G.\ D.\ Cassini may have seen it as a double star some 40 years
earlier.  Castor holds the distinction of being the first true
physical binary to be recognized as such \citep{Herschel:1803}, based
on changes in the direction of the line joining the two stars observed
over a few decades. This has been regarded by some as the first
empirical evidence that Newton's laws of gravitation apply beyond the
solar system.

The fainter star of the pair, Castor~B, was in turn discovered by
\cite{Belopolsky:1897} to be a spectroscopic binary with a period of
2.4~days, and a few years later \cite{Curtis:1906} found Castor~A to
also be a spectroscopic binary, with a longer period of 9.2~days. The
primaries of both systems are A-type stars, and the companions are M dwarfs
that are too dim to be seen spectroscopically. Their nature is
inferred from the detection of X-rays in both Castor~A and B
\citep{Schmitt:1994, Gudel:2001, Stelzer:2003}, which would be unusual
coming from stars of spectral type A, but is to be expected for M
dwarfs.  A more distant, but physically related companion to
Castor~AB, currently some 71\arcsec\ away, is also known as YY~Gem (or
Castor~C), and happens to also be a spectroscopic binary that is
double-lined and eclipsing, making this a hierarchical sextuple
system. Both of the YY~Gem components are M dwarfs as well.

Castor~A and B have yet to complete a full revolution around each
other since the first astrometric measurement was made. Their current
separation is 5\farcs5. Numerous preliminary visual orbits for the
pair have been computed over the last two centuries, with one of the latest
determinations, by \cite{DeRosa:2012}, giving a period of 467~yr and a
semimajor axis of 6\farcs8. In principle the historical velocities
can help to constrain that orbit through the difference in the
center-of-mass velocities of the two spectroscopic subsystems.
Furthermore, more than eight decades have now passed since the last
extensive sets of spectroscopic observations for both binaries, so
that additional velocities at the present time with the much higher
precision that is now possible may provide an additional constraint.
To that end, we have been monitoring Castor~A and B spectroscopically
for the past nearly 30 years.  Not only have our observations now revealed
a drift in the systemic velocities of both binaries in opposite directions,
but the sign of the velocity difference between A and B has reversed
compared to what it was a century ago, indicating they have gone through
conjunction in the outer orbit.

The dynamical masses of YY~Gem are well known from the fact that it is
eclipsing \citep{Segransan:2000, Torres:2002, Kochukhov:2019}. Those
of Castor~A and B, on the other hand, have not been determined
independently of models or other assumptions because the secondaries
have never been spatially resolved. This has been one of the main
motivations for this work.  We have pursued that challenge here
through long-baseline interferometric observations, and have succeeded
in detecting both secondaries for the first time. The combination of
those measurements, the radial velocities, the visual observations of
the outer orbit, and the Hipparcos intermediate astrometric data
should now enable the full 3D orbits to be determined, aided by
the additional constraint from the parallax of the Castor system
delivered by the Gaia mission \citep{Gaia:2016, Gaia:2022}.
The architecture of the system can therefore be
completely specified, including the true mutual inclination angles
between the three orbital planes, which are of considerable interest
for studying the dynamical evolution of the system. In addition to
holding the key to the masses of the four stars, the interferometric
observations also allow us to directly measure the absolute radii of
the primaries, providing a way to infer the age of the system using
models. All of these topics are the subject of this work, which aims
to more fully characterize the main components of Castor.

The remainder of the paper is organized as follows. In
Section~\ref{sec:observations} we describe our new interferometric and
spectroscopic observations, as well as the visual observations of
Castor~AB gathered since its discovery. We also describe the
intermediate data from the Hipparcos mission, which turn out to be very
useful as well. Our global analysis that solves simultaneously for the
orbital elements of Castor~A, Castor~B, and Castor~AB is explained in
Section~\ref{sec:analysis}, where we report the main orbital and
physical properties of the quadruple system. These results are
discussed in Section~\ref{sec:discussion}, in which we make a
determination of the age of the system using current stellar evolution
models. We also discuss there the motion of Castor~C relative to Castor~AB,
and sum up the empirical data that can constrain the dynamical
evolution of the sextuple system. Our conclusions are given in
Section~\ref{sec:conclusions}. The Appendix then gives details of
the historical radial velocity measurements, which serve to support
the accuracy of the spectroscopic orbit of the AB pair, and provide
additional information to constrain the orbit of Castor~C.

\section{Observations}
\label{sec:observations}

We begin the description of the observations for Castor with our own
interferometric and spectroscopic measurements, followed by the
extensive set of visual observations from the literature.

\subsection{Interferometry}
\label{sec:interferometry}

Interferometric observations were obtained with the CHARA Array operated by Georgia State University and located at Mount Wilson Observatory in southern California. The CHARA Array consists of six 1\,m telescopes arranged in a `Y' configuration with baselines ranging from 34\,m to 331\,m \citep{tenBrummelaar:2005}. In 2007, Castor A was observed on three nights with the original version of the Michigan InfraRed Combiner \citep[MIRC;][]{Monnier:2004, Monnier:2006}. During these observations, MIRC combined the light from the S1, E1, W1, and W2 telescopes and recorded fringes in 8 spectral channels in the $H$ band. In 2021, Castor~A and B were observed on six nights using the upgraded MIRC-X instrument \citep{Anugu:2020} using all six telescopes (S1, S2, E1, E2, W1, and W2). The two nights in March of 2021 were obtained with the $R = 190$ grism, and the remaining nights in November and December, 2021 were obtained with the $R = 50$ prism to optimize throughput for fainter targets from other
programs observed on the same nights. In November and December 2021, we obtained simultaneous $K$-band observations using the MYSTIC six-telescope combiner \citep{Monnier:2018} with the $R = 49$ prism.
We alternated between observations of unresolved calibrators stars and the science targets. The calibrators were selected using
SearchCal\footnote{\url{https://jmmc.fr/searchal}}
\citep{Chelli:2016} and are listed in Table~\ref{tab:obslog}. 

The 5\farcs5 separation between Castor A and B presented an observational challenge for the CHARA telescopes. The telescopes would sometimes get confused between the two sources and switch from locking on one component to the other. When the telescope pointing changed, we would see a change in flux on the MIRC-X detector and a corresponding disappearance of the fringes on the baselines associated with the impacted telescope. The changing amount of incoherent flux on the detector also caused some miscalibrations in the visibilities on other baselines. This problem happened more frequently in bad seeing conditions and when the telescopes were pointing to the fainter B component. 

The MIRC data were reduced and calibrated using the standard MIRC pipeline written in IDL \citep{Monnier:2007}. The MIRC-X and MYSTIC data were reduced using the standard MIRC-X pipeline (version 1.3.5) written in python\footnote{\url{https://gitlab.chara.gsu.edu/lebouquj/mircx_pipeline.git}}. On each night the calibrators were calibrated against each other to check for binarity. No evidence of binarity was
found in the calibrators based on visual inspection. The calibrated OIFITS files for Castor A and B will be available in the Optical Interferometry Database\footnote{\url{http://jmmc.fr/~webmaster/jmmc-html/oidb.htm}} and the CHARA Data Archive\footnote{\url{https://www.chara.gsu.edu/observers/database}}. 

\begin{figure}
\epsscale{1.15}
\plotone{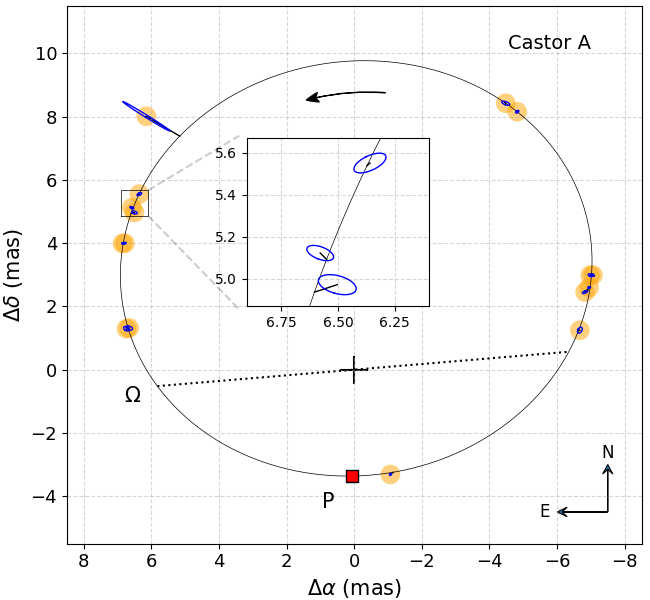}
\figcaption{CHARA observations of Castor~A along with our best-fit
  model described in Section~\ref{sec:analysis}. Except for one
  observation in the first quadrant, error ellipses and short line
  segments connecting each measurement to the predicted position on
  the orbit are generally too small to be seen. Orange circles are
  drawn at the location of each measurement for better visibility. An
  enlargement of a small section of the orbit is shown in the inset
  to illustrate the error ellipses. The
  cross at the origin marks the position of Castor~Aa, and the dotted
  line represents the line of nodes (the ascending node is marked with
  the $\Omega$ symbol).  Periastron is indicated with a red square
  labeled ``P''.\label{fig:charaA}}
\end{figure}

\begin{figure}
\epsscale{1.15}
\plotone{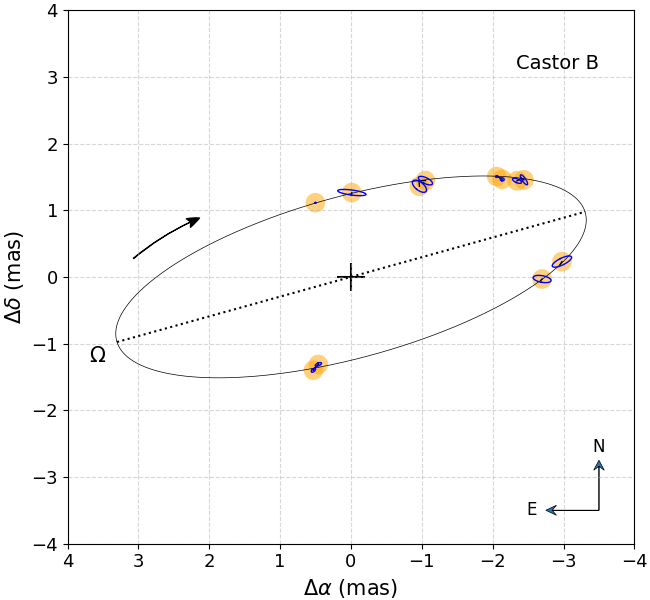}
\figcaption{Similar to Figure~\ref{fig:charaA} for Castor~B. The error
  ellipses are now visible because of the smaller scale of the
  orbit.\label{fig:charaB}}
\end{figure}

The calibrated visibilities and closure phases were fit by performing a binary grid search using software written by J.D.M.\ to solve for the binary separation $\rho$, position angle $\theta$, flux ratio, and angular diameter $\phi$ of the primary component (Aa or Ba) during each observation. The (uniform disk) angular diameters of the companions, Ab and Bb, were fixed at 0.23~mas
based on their expected sizes according to the PARSEC 1.2S stellar evolution models of \cite{Chen:2014}
for their masses as measured later.
The fit for each epoch also allowed for an over-resolved 'incoherent flux' that might arise from either seeing changes or light contamination from the far component (Castor~B if observing Castor~A, and vice versa) which might be coupled into the fiber.  For some nights with poor UV coverage, there were multiple binary positions allowable by the data; we chose the one closest to the orbital prediction.  The final values for the $H$-band flux ratios from MIRC-X are $f_{\rm Aa/Ab} = 197 \pm 12$ and $f_{\rm Ba/Bb} = 88 \pm 12$, and the corresponding $K$-band flux ratios from MYSTIC are $f_{\rm Aa/Ab} = 146 \pm 22$ and $f_{\rm Ba/Bb} = 68 \pm 10$.  The uniform disk diameters we measured with MIRC-X are $\phi_{\rm UD,Aa} = 1.273 \pm 0.003$~mas and $\phi_{\rm UD,Ba} = 1.005 \pm 0.008$~mas. These determinations were based on a global fit of the best 5 nights of data, using bootstrap sampling to estimate the errors. A similar
procedure for MYSTIC gave angular diameters of $\phi_{\rm UD,Aa} = 1.271 \pm 0.012$~mas and $\phi_{\rm UD,Ba} = 0.994 \pm 0.014$~mas. The angular diameters will be used later in Section~\ref{sec:discussion} to establish the absolute radii.
Plots of the orbits for Castor~A and B are shown in
  Figures~\ref{fig:charaA} and \ref{fig:charaB} together with our best-fit model described
  below, and the measured positions are presented in Table~\ref{tab:CHARA}. Position angles in the table are referred to the epoch of observation. For the orbital analysis below they will be corrected for precession to the year 2000. The angular separations in the table as well as the angular diameters reported above include small, empirically determined downward adjustments by factors of $1.0054 \pm 0.0006$ for MIRC-X and $1.0067 \pm 0.0007$ for MYSTIC (T.\ Gardner, priv.\ comm.), equivalent to a reduction in the respective wavelengths reported in the OIFITS files by the same factors.

\begin{deluxetable*}{llll}
\tablecaption{{CHARA MIRC, MIRC-X, and MYSTIC Observing Log} \label{tab:obslog}}
\tablehead{
\colhead{UT Date} &
\colhead{Instrument} &
\colhead{Mode} &
\colhead{Calibrators} }
\startdata
2007Nov19  & MIRC    & H-Prism50    & HD 24398 ($\zeta$ Per), HD 32630 ($\eta$ Aur) \\
2007Nov22  & MIRC   & H-Prism50     & HD 24398 ($\zeta$ Per), HD 87737 ($\eta$ Leo), HD 97633 ($\theta$ Leo) \\
2007Nov23  & MIRC   & H-Prism50     & HD 14055 ($\gamma$ Tri), HD 97633 ($\theta$ Leo) \\
2021Mar02  & MIRC-X & H-Grism190    & HD 59037, HD 71148, HD 74811 \\
2021Mar06  & MIRC-X & H-Grism190    & HD 50692, HD 59037, HD 67542 \\
2021Nov19  & MIRC-X, MYSTIC & H-Prism50, K-Prism49  & HD 50692, HD 59037, HD 71148 \\
2021Dec08  & MIRC-X, MYSTIC & H-Prism50, K-Prism49  & HD 59037, HD 67542 \\
2021Dec20  & MIRC-X, MYSTIC & H-Prism50, K-Prism49  & HD 59037, HD 67542, HD 74811 \\ [-1ex]
\enddata
\tablecomments{Calibrator diameters for the MIRC-X and MYSTIC observations were generally adopted from the JMMC Stellar Diameter Catalog \citep{Bourges:2017}: 
HD~14055 ($\theta_{\rm H} = 0.470 \pm 0.033$~mas), HD 50692 ($\theta_{\rm H} = 0.539 \pm 0.051$~mas, $\theta_{\rm K} = 0.541 \pm 0.051$~mas), HD 59037 ($\theta_{\rm H} = 0.390 \pm 0.011$~mas, $\theta_{\rm K} = 0.391 \pm 0.011$~mas), HD 67542 ($\theta_{\rm H} = 0.491 \pm 0.042$~mas, $\theta_{\rm K} = 0.493 \pm 0.042$~mas), HD 71148 ($\theta_{\rm H} = 0.462 \pm 0.011$~mas, $\theta_{\rm K} = 0.464 \pm 0.011$~mas), HD 74811 ($\theta_{\rm H} = 0.414 \pm 0.010$~mas, $\theta_{\rm K} = 0.416 \pm 0.010$~mas), HD 87737 ($\theta_{\rm H} = 0.65 \pm 0.06$~mas),
HD~97633 ($\theta_{\rm H} = 0.80 \pm 0.08$~mas),
HD~32630 \citep[$\theta_{\rm H} = 0.453 \pm 0.012$~mas;][]{Maestro:2013}, HD~24398 \citep[$\theta_{\rm H} = 0.53 \pm 0.03$~mas;][]{Challouf:2014}.}
\end{deluxetable*}

\setlength{\tabcolsep}{5pt}
\begin{deluxetable*}{lccccccclc}
\tablecaption{CHARA Measurements for Castor~A and B \label{tab:CHARA}}
\tablehead{
\colhead{BJD} &
\colhead{$\Delta t$} &
\colhead{UT Date} &
\colhead{$\rho$} &
\colhead{$\theta$} &
\colhead{$\sigma_{\rm maj}$} &
\colhead{$\sigma_{\rm min}$} &
\colhead{$\theta_{\sigma}$} &
\colhead{Instrument} &
\colhead{Orbital}
\\
\colhead{(2,400,000+)} &
\colhead{(day)} &
\colhead{} &
\colhead{(mas)} &
\colhead{(degree)} &
\colhead{(mas)} &
\colhead{(mas)} &
\colhead{(degree)} &
\colhead{} &
\colhead{Phase}
}
\startdata
\multicolumn{10}{c}{Castor A} \\ [0.5ex]
\noalign{\hrule} \\ [-1.5ex]
   54423.9754  & $-0.1121$ & 2007Nov19  &  10.100     & \phn37.56  &  0.982  &  0.045  & \phn56.29  & MIRC & 0.696 \\
   54426.9056  & $-0.1121$ & 2007Nov22  &  \phn3.482  & 197.97     &  0.043  &  0.029  & 318.61     & MIRC & 0.014 \\
   54427.9267  & $-0.1121$ & 2007Nov23  &  \phn6.791  & 280.64     &  0.119  &  0.071  & 331.70     & MIRC & 0.125 \\
   59275.6936  & $-0.0778$ & 2021Mar02  &  \phn9.475  & 329.54     &  0.066  &  0.038  & 309.61     & MIRC-X & 0.331 \\
   59275.8366  & $-0.0778$ & 2021Mar02  &  \phn9.544  & 332.11     &  0.146  &  0.058  & \phn68.51  & MIRC-X & 0.346 \\
   59279.6593  & $-0.0777$ & 2021Mar06  &  \phn8.441  & \phn49.01  &  0.091  &  0.040  & 297.55     & MIRC-X & 0.761 \\
   59279.7822  & $-0.0777$ & 2021Mar06  &  \phn8.337  & \phn52.22  &  0.073  &  0.034  & \phn67.25  & MIRC-X & 0.775 \\
   59279.8232  & $-0.0777$ & 2021Mar06  &  \phn8.186  & \phn52.73  &  0.101  &  0.050  & \phn72.92  & MIRC-X & 0.779 \\
   59538.0184  & $-0.0758$ & 2021Nov19  &  \phn7.934  & \phn59.93  &  0.029  &  0.021  & 294.62     & MIRC-X & 0.805 \\
   59538.0184  & $-0.0758$ & 2021Nov19  &  \phn7.876  & \phn59.63  &  0.047  &  0.028  & 298.39     & MYSTIC & 0.805 \\
   59557.0147  & $-0.0756$ & 2021Dec08  &  \phn6.788  & \phn78.93  &  0.123  &  0.068  & \phn66.67  & MIRC-X & 0.867 \\
   59557.0147  & $-0.0756$ & 2021Dec08  &  \phn6.862  & \phn79.26  &  0.110  &  0.071  & \phn57.61  & MYSTIC & 0.867 \\
   59568.8842  & $-0.0755$ & 2021Dec20  &  \phn7.419  & 290.63     &  0.050  &  0.032  & 283.59     & MIRC-X & 0.156 \\
   59568.8842  & $-0.0755$ & 2021Dec20  &  \phn7.258  & 289.91     &  0.102  &  0.040  & 295.79     & MYSTIC & 0.156 \\
   59568.9742  & $-0.0755$ & 2021Dec20  &  \phn7.670  & 293.08     &  0.061  &  0.032  & \phn47.94  & MIRC-X & 0.165 \\
   59568.9742  & $-0.0755$ & 2021Dec20  &  \phn7.593  & 293.29     &  0.077  &  0.041  & \phn62.14  & MYSTIC & 0.165 \\ [0.5ex]
\noalign{\hrule} \\ [-1.5ex]
\multicolumn{10}{c}{Castor~B} \\ [0.5ex]
\noalign{\hrule} \\ [-1.5ex] 
   59275.7346  &  +0.0988  & 2021Mar02  &   1.221  & \phn24.43  &  0.008  &  0.008  & \phn70.00  & MIRC-X & 0.743 \\
   59275.8056  &  +0.0988  & 2021Mar02  &   1.265  & 359.59     &  0.256  &  0.048  & \phn82.04  & MIRC-X & 0.767 \\
   59279.6873  &  +0.0987  & 2021Mar06  &   2.986  & 274.54     &  0.194  &  0.063  & 297.85     & MIRC-X & 0.092 \\
   59279.7593  &  +0.0987  & 2021Mar06  &   2.697  & 269.45     &  0.164  &  0.063  & \phn78.81  & MIRC-X & 0.117 \\
   59537.9664  &  +0.0962  & 2021Nov19  &   1.391  & 160.78     &  0.062  &  0.026  & 295.77     & MIRC-X & 0.291 \\
   59537.9664  &  +0.0962  & 2021Nov19  &   1.497  & 159.38     &  0.049  &  0.025  & 314.92     & MYSTIC & 0.291 \\
   59557.0657  &  +0.0960  & 2021Dec08  &   1.786  & 324.09     &  0.140  &  0.057  & \phn62.56  & MIRC-X & 0.813 \\
   59557.0657  &  +0.0960  & 2021Dec08  &   1.668  & 324.75     &  0.154  &  0.076  & \phn49.17  & MYSTIC & 0.813 \\
   59568.9552  &  +0.0959  & 2021Dec20  &   2.549  & 306.31     &  0.022  &  0.015  & 291.13     & MIRC-X & 0.873 \\
   59568.9552  &  +0.0959  & 2021Dec20  &   2.587  & 304.59     &  0.036  &  0.029  & \phn75.19  & MYSTIC & 0.873 \\
   59569.0202  &  +0.0959  & 2021Dec20  &   2.843  & 300.95     &  0.111  &  0.033  & \phn32.15  & MIRC-X & 0.895 \\
   59569.0202  &  +0.0959  & 2021Dec20  &   2.752  & 301.73     &  0.088  &  0.036  & \phn65.23  & MYSTIC & 0.895 \\ [-1ex]
\enddata
\tablecomments{Column 2 ($\Delta t$) is the light travel time correction applied
to the observed times in the first column during our analysis to reduce them to the center of mass
of the quadruple system. Position angles $\theta$ are referred to the epoch of observation. The symbols $\sigma_{\rm maj}$,
$\sigma_{\rm min}$, and $\theta_{\sigma}$ represent the major and minor axes of the
formal error ellipse for each measurement, and the position angle of the major axis at the epoch of observation, measured in the
usual direction from north to east. Observations with MIRC and MIRC-X were made in the $H$ band,
and those with MYSTIC in the $K$ band. Orbital phases in each
orbit are computed from the ephemerides given in Section~\ref{sec:analysis}.}
\end{deluxetable*}
\setlength{\tabcolsep}{6pt}

\subsection{Spectroscopy}
\label{sec:spectroscopy}

Observations of Castor~A and B at the Center for Astrophysics (CfA)
began in January of 1993. They were made with two nearly identical
copies of the Digital Speedometer \citep[DS;][]{Latham:1992} on the
1.5\,m Wyeth reflector at the (now closed) Oak Ridge Observatory in the
town of Harvard (MA), and on the 1.5\,m Tillinghast reflector at the
Fred L.\ Whipple Observatory on Mount Hopkins (AZ). These echelle
instruments used intensified Reticon detectors that recorded a single
order 45~\AA\ wide centered on the \ion{Mg}{1}~b triplet near
5187~\AA, with a resolving power of $R \approx 35,000$. A total of 67
spectra were obtained for Castor~A with signal-to-noise ratios at
5187~\AA\ ranging from 30 to 116 per resolution element of 8.5~\kms.
However, at the higher levels the limitation is systematic errors in
the flatfield corrections rather than photon noise. For Castor~B we
gathered 65 spectra, and the signal-to-noise ratios are 28--118. An
additional six observations were made of the combined light of
Castor~A and B, when they could not be separated under poor seeing
conditions. Those have signal-to-noise ratios of 51--86. The last of
the observations with these instruments were gathered in May of 2009.

Wavelength solutions relied on exposures of a thorium-argon lamp taken
before and after each science exposure, and observations of the
morning and evening sky were used to monitor the velocity zero
point. Small run-to-run corrections generally under 2~\kms\ were
applied to the radial velocities described below
\citep[see][]{Latham:1992}, placing the measurements from both
telescopes on the same system. This system is slightly offset from the
IAU reference frame by 0.14~\kms\ \citep{Stefanik:1999}, as determined
from observations of minor planets in the solar system. In order to
remove this shift, we added a correction of +0.14~\kms\ to the raw
velocities. By construction our velocities have the gravitational
redshift and convective blueshift of the Sun subtracted out (see the
Appendix).

Starting in October of 2009, spectroscopic monitoring was continued
with the Tillinghast Reflector Echelle Spectrograph
\citep[TRES;][]{Furesz:2008, Szentgyorgyi:2007}, which is a modern
bench-mounted, fiber-fed instrument on the 1.5\,m telescope in Arizona.
The resolving power is $R \approx 44,000$, and a CCD records 51 orders
over the 3800--9100~\AA\ range. We collected 174 and 164 observations
of Castor~A and B, respectively, through April of 2022. The
signal-to-noise ratios at 5187~\AA\ range from 91 to 1383 for
Castor~A, and 78 to 1015 for Castor~B per resolution element of
6.8~\kms. Thorium-argon exposures were
used as before for the wavelength solutions, and changes in the
velocity zero point were monitored with observations of IAU standard
stars. Asteroid observations were then employed to transfer the raw
TRES velocities to an absolute system, as done for the DS instruments.

Radial velocities from all spectra were derived by cross-correlation
against synthetic templates taken from a large pre-computed library based on
model atmospheres by R.\ L.\ Kurucz, and a line list tuned to better
match real stars \citep[see][]{Nordstrom:1994, Latham:2002}. These
templates cover a limited wavelength region near the Mg triplet.  A
complication in this case is that the composition of both Castor~A and
Castor~B is anomalous. They are classified as metallic-line A stars
\citep[see, e.g.,][]{Conti:1965, Smith:1974, Roby:1990}. The
abundances of the iron-peak elements are enhanced, while others such
as Ca are depleted in Castor~B, but enhanced in Castor~A, which is one
of the hotter Am stars \citep{Smith:1974}. Other elements tend to
follow the typical abundance pattern for Am stars, although with some
other differences particularly for Castor~A.  Consequently, synthetic
spectra with solar-scaled abundances such as ours will not match the
real stars as well as they could at any metallicity, and as a result
the velocity precision may suffer to some degree. More importantly,
systematic errors in the velocities may be introduced such that it
becomes difficult to place them accurately on a well-defined absolute
zero point to much better than $\sim$1~\kms, as we had intended.  To
attempt to compensate for the peculiar abundances, we adopted a
supersolar metallicity for Castor~B (${\rm [Fe/H]} = +0.5$), the star
in which the anomalies appear more pronounced. Solar composition was
adopted for Castor~A. Effective temperatures and rotational
broadenings for the templates were taken to be 9750~K and 20~\kms\ for
Castor~A, and 8250~K and 30~\kms\ for Castor~B, which provided the
best match to the observed spectra of each star. The surface gravities
were held fixed at $\log g = 4.0$.

The velocities for Castor~A and B from both instruments are listed in
Tables~\ref{tab:rvA} and \ref{tab:rvB}, respectively, along with their
uncertainties. Typically uncertainties are about 0.9 and 1.2~\kms\ for
the DS measurements of Castor~A and B, and about 0.07 and
0.06~\kms\ for the TRES measurements.

\setlength{\tabcolsep}{4pt}
\begin{deluxetable}{lccccc}
\tablecaption{Radial Velocity Measurements for Castor~A \label{tab:rvA}}
\tablehead{
\colhead{HJD} &
\colhead{$\Delta t$} &
\colhead{$RV_{\rm Aa}$} &
\colhead{$\sigma_{\rm RV}$} &
\colhead{Inner} &
\colhead{Outer}
\\
\colhead{(2,400,000+)} &
\colhead{(day)} &
\colhead{(\kms)} &
\colhead{(\kms)} &
\colhead{Phase} &
\colhead{Phase}
}
\startdata
 49004.7813  & $-$0.1417  &    \phn2.00      &  1.50  &  0.465  &  0.073 \\
 49018.7363  & $-$0.1417  & \phn$-$8.14\phs  &  3.36  &  0.980  &  0.073 \\
 51265.6006  & $-$0.1309  &    $-$12.19\phs  &  0.95  &  0.868  &  0.086 \\
 52293.7003  & $-$0.1252  &    \phn2.44      &  1.31  &  0.464  &  0.092 \\
 52293.7110  & $-$0.1252  &    \phn1.03      &  0.79  &  0.465  &  0.092 \\ [-1ex]
\enddata
\tablecomments{Column~2 lists the light travel time correction in the
  outer orbit.
  Observations prior to HJD $2,\!455,\!000$ were made with the DS
  instrument, and more recent ones with TRES. The uncertainties have
  been adjusted as explained in Section~\ref{sec:analysis} to be more
  realistic.  Orbital phases in the last column are computed using the
  ephemerides given later in Section~\ref{sec:analysis}. (This table is
  available in its entirety in machine-readable form.)}
\end{deluxetable}
\setlength{\tabcolsep}{6pt}

\setlength{\tabcolsep}{4pt}
\begin{deluxetable}{lccccc}
\tablecaption{Radial Velocity Measurements for Castor~B \label{tab:rvB}}
\tablehead{
\colhead{HJD} &
\colhead{$\Delta t$} &
\colhead{$RV_{\rm Ba}$} &
\colhead{$\sigma_{\rm RV}$} &
\colhead{Inner} &
\colhead{Outer}
\\
\colhead{(2,400,000+)} &
\colhead{(day)} &
\colhead{(\kms)} &
\colhead{(\kms)} &
\colhead{Phase} &
\colhead{Phase}
}
\startdata
 49028.7560  &  +0.1795  &  $-$25.90\phs  &  3.38  &  0.539  &  0.073 \\
 51292.5907  &  +0.1657  &  $-$17.91\phs  &  1.38  &  0.609  &  0.087 \\
 52293.7003  &  +0.1587  &  $-$25.10\phs  &  2.11  &  0.475  &  0.092 \\
 52293.7058  &  +0.1587  &  $-$27.77\phs  &  1.54  &  0.477  &  0.092 \\
 52309.7103  &  +0.1586  &     33.76      &  1.08  &  0.942  &  0.093 \\ [-1ex]
\enddata
\tablecomments{Column~2 lists the light travel time correction in the
  outer orbit.
  Observations prior to HJD $2,\!455,\!000$ were made with the DS
  instrument, and more recent ones with TRES. The uncertainties have
  been adjusted as explained in Section~\ref{sec:analysis} to be more
  realistic.  Orbital phases in the last column are computed using the
  ephemerides given later in Section~\ref{sec:analysis}. (This table is
  available in its entirety in machine-readable form.)}
\end{deluxetable}
\setlength{\tabcolsep}{6pt}

\begin{figure}
\epsscale{1.15}
\plotone{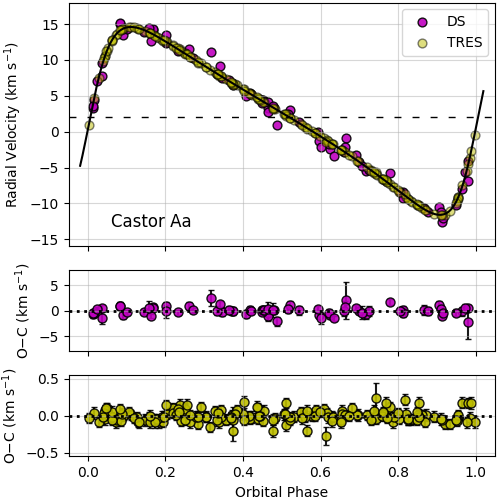}
\figcaption{Radial velocity measurements for Castor~Aa from the DS and
  TRES instruments. The solid curve is our model from
  Section~\ref{sec:analysis}, and motion in the outer orbit has been
  subtracted from the measurements for display purposes.  The dashed
  line marks the center-of-mass velocity of the quadruple system.
  Errorbars are not shown in the top panel for clarity. The lower
  panels display the residuals separately for the two instruments
  (note the different scales). \label{fig:RVinA}}
\end{figure}

\begin{figure}
\epsscale{1.15}
\plotone{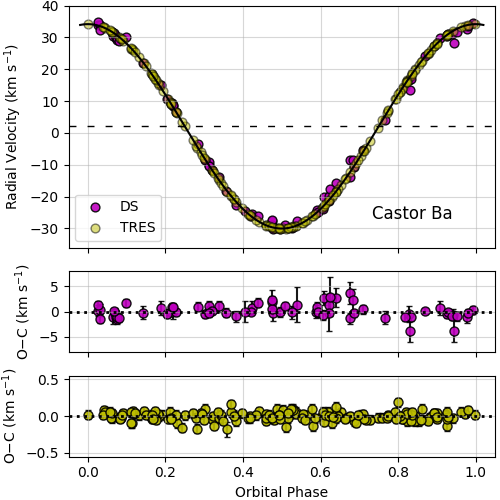}
\figcaption{Same as Figure~\ref{fig:RVinA} for
  Castor~Ba. \label{fig:RVinB}}
\end{figure}

Aside from the six DS spectra of the combined light mentioned earlier,
close examination showed that several other spectra from the DS and
TRES had contamination from the other component of the visual pair.
In these cases the velocities for the dominant star were derived using
TODCOR \citep{Zucker:1994}, which is a two-dimensional
cross-correlation technique. All velocities from both instruments are
shown in Figure~\ref{fig:RVinA} for Castor~A, and in
Figure~\ref{fig:RVinB} for Castor~B, along with our best orbit model
described later.

As indicated in Section~\ref{sec:introduction}, radial velocity
measurements of both Castor~A and B have been collected by many
observers for more than a century. While the zero points of those
historical observations are not always well defined, the measurements are still
useful as a check on our global solution because they sample a part of
the outer AB orbit that happens to correspond to maximum velocity
separation between the components. Details of these observations are
provided in the Appendix with a discussion of their use for our
purposes, and a comparison with the results of our analysis is given
later in Section~\ref{sec:analysis}.

\subsection{Visual observations}
\label{sec:visual}

Measurements of the relative position between Castor~A and B have been
made fairly regularly starting a few decades after the discovery of
its binary nature in 1718. Castor ranks among the ten visual binaries
with the most measurements recorded. It has more than 1500 entries in
the regularly updated Washington Double Star Catalog
\citep{Worley:1997, Mason:2001}, many being averages over up to 20
separate nights. These observations were all kindly provided by
R.\ Matson (USNO), with the most recent one being from 2020. We have
supplemented these measurements here with a few others gathered mostly
from the early literature \citep{Herschel:1803, Herschel:1824,
  Herschel:1833}. The vast majority of the observations ($\sim$1100)
have been made with a filar micrometer, and others were gathered with
photographic, speckle interferometry, or other measuring techniques.

As measurement uncertainties for this type of observation have
typically not been published, especially for the older data, we have
adopted the general scheme described by \cite{Douglass:1992} that
assigns errors according to the telescope aperture and measuring
technique. Additionally, we have chosen to double the uncertainties
for observations prior to 1830, which tend to show more scatter. As
with the CHARA observations, all position angle measurements have been
corrected for precession to the year 2000.

\begin{figure}[b]
\begin{tabular}{cc}
{\hspace{-5pt}\includegraphics[height=8.0cm]{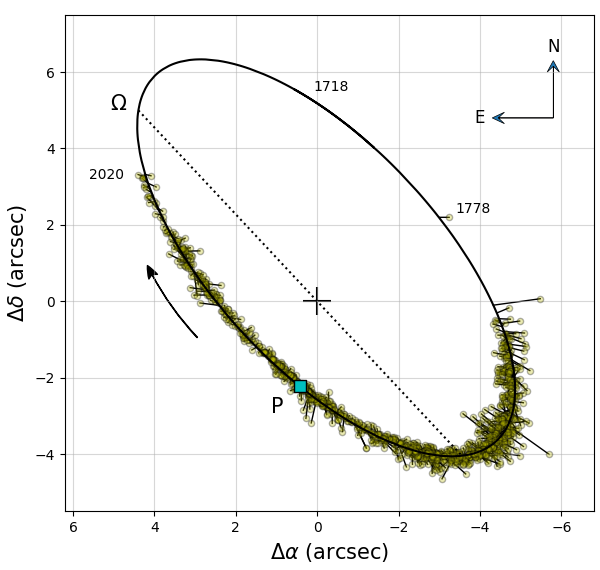}} \\
{\hspace{-3pt}\includegraphics[height=7.2cm]{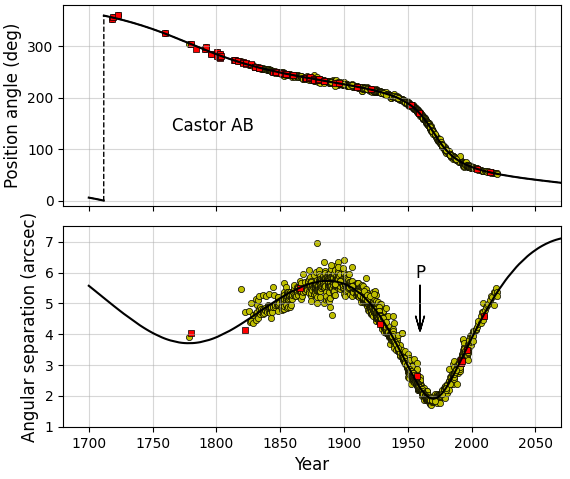}} \\
\end{tabular}

\figcaption{\emph{Top}: Measurements of the relative position of
  Castor~AB from the Washington Double Star catalog
  \citep{Worley:1997, Mason:2001} along with our best-fit model
  described in Section~\ref{sec:analysis}. 
  Thin line segments connect the measured position to the expected location
  from our model. The dotted line represents
  the line of nodes (the $\Omega$ symbol indicates the ascending
  node), and periastron is marked with a cyan square (``P''). Only
  measurements that have both a position angle and a separation are
  shown. Others are shown below. \emph{Bottom panels}: All position
  angle and separation measurements as a function of time. Incomplete observations
  missing either the position angle or the separation are distinguished
  with red squares. The arrow
  marks the time of periastron passage.\label{fig:visual}}

\end{figure}

The visual observations now cover two thirds of the AB orbit. They are
shown in the top panel of Figure~\ref{fig:visual} along with our orbit model
described below. Of the 1507 distinct epochs of observation, several
dozen of them are missing either the position angle or,
more commonly, the angular separation. While this does not
preclude their use for the orbital analysis, it does prevent them from
being shown in the figure.  In particular, the early position angle
measurements recorded for six decades before the first complete
observation by Herschel in 1778\footnote{Herschel's angular separation
  measurements up to about 1825 do not correspond to the distance
  between the star centers, but instead represent the separation
  between the outer edges of the apparent disks of the stars as seen
  with his telescope. They therefore included the sum of the apparent
  semidiameters of both components, as Herschel himself pointed out.
  An estimate of this excess, 1\farcs24 at the typical power he used
  at the telescope, follows from his detailed description of his
  measuring procedure \citep{Herschel:1803}, and we have subtracted
  this amount from his measurements of $\rho$ before 1825 to remove
  the bias. With this adjustment, we find that all his values agree
  very well with predictions from an orbit computed without them. In
  later years Herschel used a micrometer of a different design, and
  reported proper center-to-center angular separations.} are quite
valuable for constraining the period. To show those observations, we
represent them graphically as a function of time in the lower panels
of Figure~\ref{fig:visual}.

\subsection{Hipparcos observations}
\label{sec:hipparcos}

Additional astrometry is available from the Hipparcos mission
\citep{ESA:1997}, which gathered measurements of Castor (HIP~36850)
over a period of three years between March of 1990 and March of 1993.
The AB pair was easily resolved, and an estimate was reported of its
relative position ($\rho = 3\farcs12$, $\theta = 75\fdg5$) along with
the proper motion components and trigonometric parallax of the system.
However, orbital motion was not accounted for in deriving those
quantities, and we estimate the change in the relative position
amounted to approximately 350~mas over that interval, which is not
negligible compared to the typical few milli-arc second measurement
precision of the satellite. The published results are therefore likely
biased to some degree. The revised Hipparcos catalog resulting from a
re-reduction of the original mission data \citep{vanLeeuwen:2007}
suffers from the same drawback.

In order to fully exploit these observations and extract useful
constraints on the orbit of the binary, we have made use here of the
Hipparcos intermediate astrometric data (IAD), which permit a reanalysis of
the satellite observations for an individual object with arbitrarily
complex models. The IAD come in two flavors: one-dimensional abscissa residuals
(in units of mas), and transit data. For this work we used the transit data
\citep[TD;][]{Quist:1999}, which are publicly available for most binary or multiple
systems and are given as five Fourier coefficients $b_1 \ldots b_5$ 
describing the signal produced at each epoch by an object as it crosses
(or ``transits'') the focal grid.  An advantage of the transit data over
the abscissa residuals is
that they allow the extraction of photometric information as well.

\section{Orbital analysis}
\label{sec:analysis}

Our approach to determining the orbital elements of Castor~A,
Castor~B, and Castor~AB is to use all the astrometric and
spectroscopic observations together, and to solve simultaneously for
the elements of the three orbits. We will assume for this work that
the motion in each subsystem is purely Keplerian, i.e., that it is
unperturbed by the other bodies in this hierarchical quadruple system.

The elements of the 9.2~day orbit of Castor~A are the period ($P_{\rm
  A}$), the angular semimajor axis ($a^{\prime\prime}_{\rm A}$), the
eccentricity ($e_{\rm A}$), the argument of periastron for the primary
star Aa ($\omega_{\rm Aa}$), the inclination angle to the line of
sight ($i_{\rm A}$), the position angle of the ascending node for the
year 2000 ($\Omega_{\rm A}$), a reference time of periastron passage
($T_{\rm A}$), and the velocity semiamplitude of the primary ($K_{\rm
  Aa}$). The orbit of Castor~B is described by a similar set of
elements. The velocity semiamplitudes of the secondaries in the two
systems, $K_{\rm Ab}$ and $K_{\rm Bb}$, are not directly measurable
because the stars are not detected spectroscopically.

The outer orbit of Castor~AB is represented by the following elements:
the period ($P_{\rm AB}$), the angular semimajor axis
($a^{\prime\prime}_{\rm AB}$), the eccentricity and argument of
periastron for the secondary ($e_{\rm AB}$, $\omega_{\rm B}$), the
inclination angle ($i_{\rm AB}$), the position angle of the ascending
node at epoch 2000 ($\Omega_{\rm AB}$), the time of periastron passage
($T_{\rm AB}$), the velocity semiamplitudes ($K_{\rm A}$ and $K_{\rm
  B}$), and the center-of-mass velocity of the quadruple system
($\gamma_{\rm AB}$).  However, by virtue of Kepler's third law there
are redundancies such that two of these elements can be obtained from
a combination of others:
\begin{equation}
a^{\prime\prime}_{\rm AB} = \left[ P_{\rm AB}^2 \left(
\frac{a^{\prime\prime\,3}_{\rm A}}{P_{\rm A}^2} +
\frac{a^{\prime\prime\,3}_{\rm B}}{P_{\rm B}^2}
\right) \right]^{1/3},~ \\
K_{\rm B} = K_{\rm A} \frac{a^{\prime\prime\,3}_{\rm A}/P_{\rm A}^2}
{a^{\prime\prime\,3}_{\rm B}/P_{\rm B}^2}~.
\label{eq:1}
\end{equation}
These two elements can therefore be eliminated as adjustable variables.

While the interferometric observations constrain the astrometric orbits of both inner binaries, the individual masses of their components cannot be obtained directly because both Castor~A and Castor~B show only the lines of the primaries in their spectra. The only mass information they provide comes in the form of the so-called mass function: $f(M) = (M_{\rm Ab} \sin i_{\rm A})^3 / (M_{\rm Aa} + M_{\rm Ab})^2$ for Castor~A, and an analogous expression for Castor~B. One additional piece of information is needed to infer the individual masses, such as a parallax for the system. The parallax (distance) allows the total masses $M_{\rm A} = M_{\rm Aa} + M_{\rm Ab}$ and $M_{\rm B} = M_{\rm Ba} + M_{\rm Bb}$ to be determined from Kepler's third law because the angular semimajor axes from CHARA and the orbital periods are also known. As the inclination angles $i_{\rm A}$ and $i_{\rm B}$ are determined as well, use of $f(M)$ then gives us access to the individual masses.

The outer orbit is
effectively double-lined because we measure the velocities of both A
and B (and can subtract the known motion in each of these inner orbits, leaving only the motion in the outer orbit), so the combination of the spectroscopic and astrometric elements
of the wide orbit allows the orbital parallax to be determined:
\begin{equation}
\pi_{\rm orb} = \frac{2\pi}{P_{\rm AB}} \frac{ a^{\prime\prime}_{\rm AB} \sin i_{\rm AB} }
{\sqrt{1-e_{\rm AB}^2}(K_{\rm A}+K_{\rm B})}~,
\label{eq:2}
\end{equation}
where $a^{\prime\prime}_{\rm AB}$ and $K_{\rm B}$ follow from
eq.(\ref{eq:1}). This provides the missing piece of information mentioned above.

In practice, however, $\pi_{\rm orb}$ is not very well determined because
the constraint on the velocity semiamplitude $K_{\rm A}$ is weak due
to the fact that our spectroscopic observations cover only a short
segment ($< 10$\%) of the orbit. As it turns out, the
parallax of Castor has always been rather poorly determined. The
ground-based measurements as summarized in the Yale Parallax Catalog
\citep{vanAltena:1995} give a weighted average of $74.7 \pm 2.5$~mas
based on 16 independent estimates, some of which are rather discordant
on account of the brightness and closeness of the pair. For details on
those parallax determinations, see \cite{Torres:2002}. The original
Hipparcos catalog \citep{ESA:1997} reported a considerably different
value of $63.27 \pm 1.23$~mas, whereas the 2007 re-reduction
\citep{vanLeeuwen:2007} gave a similar result but with a much larger
uncertainty: $64.12 \pm 3.75$~mas. Neither of these space-based
determinations accounted for orbital motion. Castor is so bright that
the entry in the current third data release (DR3) from the Gaia
mission \citep{Gaia:2022} does not include a value for its parallax,
and has very little other information aside from the position.

By incorporating the Hipparcos intermediate data into our solution we
gain a new handle on the parallax, as well as some constraint on the
AB orbit. This introduces several new adjustable parameters in our
analysis, which are: corrections $\Delta\alpha^*$ and $\Delta\delta$ to
the catalog values of the position of the barycenter at the mean
catalog epoch of 1991.25, corrections $\Delta\mu_{\alpha}^*$ and
$\Delta\mu_{\delta}$ to the proper motion components\footnote{We
  follow here the practice in the Hipparcos catalog of defining
  $\Delta\alpha^* \equiv \Delta\alpha \cos\delta$ and
  $\Delta\mu_{\alpha}^* = \Delta\mu_{\alpha} \cos\delta$.}, and the
apparent magnitudes of stars Aa and Ba in the $Hp$ bandpass of the
satellite, $Hp_{\rm Aa}$ and $Hp_{\rm Ba}$. The companions, Ab and Bb,
are assumed to have negligible light. The formalism for incorporating
the Hipparcos transit data in an orbital solution is described by
\cite{Quist:1999} and in the original Hipparcos documentation. In
addition to the motion in the AB orbit as well as parallactic and
proper motion, our model for the Hipparcos data accounted for the
short-period wobble of stars Aa and Ba in their respective inner
orbits.

By far the strongest constraint on the parallax comes from the Gaia
mission. Even though, as mentioned above, there is no value for
Castor~AB itself, Gaia does report a highly precise parallax for
Castor~C (YY~Gem), which is a much fainter physically bound companion.
We use this value, $\pi_{\rm Gaia} = 66.350 \pm 0.036$~mas, as an
independent measurement, along with its uncertainty.\footnote{This
  value includes a zero point correction of +0.039~mas that we have applied
  following \cite{Lindegren:2021}. Also, as recommended by \cite{ElBadry:2021},
  we have increased the nominal Gaia uncertainty of 0.024~mas by a
  factor of 1.13, and to be conservative, we have further
  increased it to account for a possible
  difference in distance compared to Castor~AB. The linear semimajor
  axis of Castor~C around Castor~AB is roughly 1100~au
  ($\sim$71\arcsec\ separation at $\sim$15~pc). If entirely along the
  line of sight, this represents a fraction 0.036\% of the distance,
  corresponding to a 0.024~mas difference in the parallax. We added this
  amount in quadrature to the Gaia uncertainty, to obtain a final error
  of 0.036~mas.} This then enables
the masses of all four stars in Castor~AB to be determined, as explained above. Indirectly
it also helps to define the scale of the Castor~AB orbit, because once
the masses and the parallax are known, the ratio
$a^{\prime\prime\,3}_{\rm AB}/P_{\rm AB}^2$ follows from Kepler's
third law.

To account for possible differences in the zero points of our velocity
measurements for Castor~A and B from the DS and TRES instruments as
discussed in Section~\ref{sec:spectroscopy}, we introduced three
additional adjustable variables. They represent offsets relative to the
DS velocities of Castor~A: $\Delta_{\rm DS,B}$ for the DS velocities
of Castor~B, $\Delta_{\rm TRES,A}$ for the TRES velocities of
Castor~A, and $\Delta_{\rm TRES,B}$ for the TRES velocities of
Castor~B.

We solved for the 33 adjustable parameters simultaneously by
non-linear least squares \citep[see, e.g.,][]{Press:1992}. In many
cases the formal uncertainties for the different types of measurements
may be underestimated, and in some they may be overestimated. In order to
apply the proper weights for our global fit, we scaled the formal
uncertainties by iterations in order to achieve reduced chi-squared
values near unity, separately for each kind of observation.  The orbit
of Castor~B was assumed to be circular (effectively leaving 31 adjustable
parameters), as initial solutions with the eccentricity free indicated
$e_{\rm B}$ is negligible. Rather than a time of periastron passage, in this
case $T_{\rm B}$ corresponds to a reference time of maximum velocity
for star Ba. Results are reported in Table~\ref{tab:results}. Our
solution accounts for light travel time for the inner binaries due
to their changing distance from the observer, and the
reference times for both subsystems ($T_{\rm A}$ and $T_{\rm B}$)
are given reduced to the barycenter
of the quadruple system \citep[see][]{Irwin:1952, Irwin:1959}. The
corrections to the times of observation of the radial velocities range from $-0.07$ to
$-0.14$~days for Castor~A, and from $+0.09$ to $+0.18$~days for
Castor~B. For the latter this is a non-negligible fraction of an
orbital cycle ($\sim$6\%).

\setlength{\tabcolsep}{10pt}
\begin{deluxetable}{lc}
\tablecaption{Results From Our Global Orbital Solution\label{tab:results}}
\tablehead{ \colhead{~~~~~~~~~~~Parameter~~~~~~~~~~~} & \colhead{Value}}
\startdata
\multicolumn{2}{c}{Outer orbit (Castor AB)} \\ [1ex]
\noalign{\hrule} \\ [-1.5ex]
 $P_{\rm AB}$ (year)               &  $459.1 \pm 2.3$             \\ 
 $e_{\rm AB}$                      &  $0.3382 \pm 0.0023$         \\
 $i_{\rm AB}$ (degree)             &  $115.107 \pm 0.060$         \\
 $\omega_{\rm B}$ (degree)         &  $251.84 \pm 0.38$           \\ 
 $\Omega_{\rm AB}$ (degree)        &  $41.304 \pm 0.085$          \\ 
 $T_{\rm AB}$ (year)\tablenotemark{a}  &  $1959.59 \pm 0.21$          \\ 
 $\gamma_{\rm AB}$ (\kms)\tablenotemark{b}  &  $+2.057 \pm 0.084$            \\ 
 $K_{\rm A}$ (\kms)                &  $2.789 \pm 0.021$           \\ 
 $\Delta_{\rm DS,B}$ (\kms)        &  $-0.11 \pm 0.14$            \\
 $\Delta_{\rm TRES,A}$ (\kms)      &  $-0.637 \pm 0.082$            \\
 $\Delta_{\rm TRES,B}$ (\kms)      &  $+0.269 \pm 0.088$            \\ [1ex]
\noalign{\hrule} \\ [-1.5ex]
\multicolumn{2}{c}{Castor~A} \\ [1ex]
\noalign{\hrule} \\ [-1.5ex]
 $P_{\rm A}$ (day)                 &  $9.2127496 \pm 0.0000052$   \\ 
 $a^{\prime\prime}_{\rm A}$ (mas)  &  $8.002 \pm 0.014$           \\
 $e_{\rm A}$                       &  $0.48769 \pm 0.00048$       \\
 $i_{\rm A}$ (degree)              &  $35.00 \pm 0.24$            \\
 $\omega_{\rm Aa}$ (degree)        &  $264.968 \pm 0.085$         \\ 
 $\Omega_{\rm A}$ (degree)         &  $95.100 \pm 0.093$          \\ 
 $T_{\rm A}$ (HJD$-$2,400,000)\tablenotemark{a}  &  $55817.7868 \pm 0.0018$     \\ 
 $K_{\rm Aa}$ (\kms)               &  $13.0933 \pm 0.0092$        \\ [1ex]
\noalign{\hrule} \\ [-1.5ex]
\multicolumn{2}{c}{Castor~B} \\ [1ex]
\noalign{\hrule} \\ [-1.5ex]
 $P_{\rm B}$ (day)                 &  $2.92835083 \pm 0.00000031$ \\ 
 $a^{\prime\prime}_{\rm B}$ (mas)  &  $3.4442 \pm 0.0093$         \\
 $i_{\rm B}$ (degree)              &  $110.50 \pm 0.12$           \\
 $\Omega_{\rm B}$ (degree)         &  $106.47 \pm 0.19$           \\ 
 $T_{\rm B}$ (HJD$-$2,400,000)\tablenotemark{a}  &  $56705.4942 \pm 0.0012$     \\ 
 $K_{\rm Ba}$ (\kms)               &  $32.0921 \pm 0.0064$        \\ [1ex]
\noalign{\hrule} \\ [-1.5ex]
\multicolumn{2}{c}{Hipparcos parameters for Castor AB} \\ [1ex]
\noalign{\hrule} \\ [-1.5ex]
 $\Delta\alpha^*$ (mas)            &  $+10.5 \pm 8.9$              \\
 $\Delta\delta$ (mas)              &  $-18.8 \pm 2.3$             \\
 $\Delta\mu_{\alpha}^*$ (mas yr$^{-1}$)  &  $-3.67 \pm 0.46$      \\
 $\Delta\mu_{\delta}$ (mas yr$^{-1}$)    &  $-1.20 \pm 0.43$      \\
 $Hp_{\rm A}$ (mag)                &  $1.9342 \pm 0.0010$         \\
 $Hp_{\rm B}$ (mag)                &  $2.9740 \pm 0.0026$ \\ [-1ex]
\enddata

\tablecomments{Multiplicative scale factors applied to the formal
  measurement uncertainties to reach reduced $\chi^2$ values near
  unity are as follows. For $\theta$ and $\rho$: 0.96 and 1.14; for the DS
  velocities of Castor~A and B: 1.24 and 1.17; for the TRES velocities: 1.11
  and 1.02; for the CHARA observations of Castor~A and B: 0.85 and 0.79;
  and for the Hipparcos coefficients $b_1 \ldots b_5$: 9.8, 4.3, 3.5, 4.6,
  7.8. The number of observations used of each kind are: $\theta$
  (1498), $\rho$ (1450), DS (73 and 71 for Castor~A and B), TRES (174
  and 164), CHARA (16 and 12 [$\theta$, $\rho$] pairs), Hipparcos ($57
  \times 5$ Fourier coefficients $b_i$), and Gaia parallax (1).}

\tablenotetext{a}{$T_{\rm AB}$ and $T_{\rm A}$ are reference times of
  periastron passage. $T_{\rm B}$ is a reference time of maximum
  primary velocity. All times are referred to the barycenter of the
  quadruple system.}

\tablenotetext{b}{The uncertainty reflects only the statistical
  error. Systematic errors due to template mismatch caused by the Am
  nature of the stars is difficult to quantify (see
  Section~\ref{sec:spectroscopy}).}

\end{deluxetable}

Other orbital and physical properties derived from the orbital
elements are presented in
Table~\ref{tab:derived}, and include the semimajor axis of the outer
orbit, the parallax, the masses and mass ratios, the velocity
semiamplitudes inferred for the secondaries in the inner orbits, and
the true angles $\phi$ between the three orbital planes. The angle
between the inner orbits of Castor~A and B was calculated with the
expression
\begin{equation}
\cos\phi_{\rm A,B} = \cos i_{\rm A} \cos i_{\rm B} + \sin i_{\rm A} \sin i_{\rm B}
\cos(\Omega_{\rm A}-\Omega_{\rm B})~,
\end{equation}
and similar expressions hold for $\phi_{\rm A,AB}$ and $\phi_{\rm B,AB}$.

Most of the elements from our solution of the wide orbit are quite
similar to those of the analysis by \cite{DeRosa:2012}, and agree
within their combined uncertainties. Exceptions are the inclination
angle (14$\sigma$ difference), the argument of periastron
(3.6$\sigma$), and the reference time of periastron passage
(6.6$\sigma$). All our uncertainties are smaller than theirs.

We note that the astrometric coverage of the Castor~B orbit from CHARA
is somewhat incomplete, as it lacks observations in the eastern
portion of the orbit (see Figure~\ref{fig:charaB}). There are also
fewer observations for Castor~B than for Castor~A. To assess the
extent to which this might affect the robustness of the orbit of B, we
carried out two experiments. In the first we removed one observation
at a time and repeated the global orbital fit. The results for all
parameters were always within 1$\sigma$ of our adopted values in
Table~\ref{tab:results}. This is a consequence of the fact that the orbit is
constrained not only by the CHARA observations themselves but also by
the numerous radial velocity measurements, which provide a strong
handle on most of the elements. In the second experiment we randomly
perturbed all of the observations within their respective error
ellipses, and carried out a new global solution. We repeated this 500
times. In all instances the elements were well within their formal
uncertainties, except for the inclination angle and $\Omega_{\rm B}$,
which deviated by up to about 2$\sigma$ from the values in
Table~\ref{tab:results}.  However, the most meaningful properties of
Castor~B for our purposes in the following section, i.e., the component masses,
varied by less than 1$\sigma$, supporting our conclusion that their
determination is robust.

\setlength{\tabcolsep}{10pt}
\begin{deluxetable}{lc}
\tablecaption{Derived Properties of the Castor System
\label{tab:derived}}
\tablehead{\colhead{~~~~~~~~~~~Parameter~~~~~~~~~~~} & \colhead{Value}}
\startdata
 $a^{\prime\prime}_{\rm AB}$ (\arcsec)      &  $6.722 \pm 0.021$      \\ 
 $\pi_{\rm orb}$ (mas)                      &  $66.356 \pm 0.041$     \\
 Distance (pc)                              &  $15.0703 \pm 0.0082$   \\ [1ex]
 $M_{\rm AB}$ ($M_{\sun}$)                  &  $4.933 \pm 0.016$      \\ [1ex]
 $M_{\rm A}$ ($M_{\sun}$)                   &  $2.757 \pm 0.015$      \\
 $M_{\rm Aa}$ ($M_{\sun}$)                  &  $2.371 \pm 0.015$      \\
 $M_{\rm Ab}$ ($M_{\sun}$)                  &  $0.3859 \pm 0.0018$    \\ [1ex]
 $M_{\rm B}$ ($M_{\sun}$)                   &  $2.176 \pm 0.018$      \\
 $M_{\rm Ba}$ ($M_{\sun}$)                  &  $1.789 \pm 0.016$      \\
 $M_{\rm Bb}$ ($M_{\sun}$)                  &  $0.3865 \pm 0.0020$    \\ [1ex]
 $q_{\rm AB} \equiv M_{\rm B}/M_{\rm A}$    &  $0.7891 \pm 0.0094$    \\
 $q_{\rm A} \equiv M_{\rm Ab}/M_{\rm Aa}$   &  $0.1627 \pm 0.0014$    \\
 $q_{\rm B} \equiv M_{\rm Bb}/M_{\rm Ba}$   &  $0.21606 \pm 0.00084$  \\ [1ex]
 $K_{\rm B}$ (\kms)                         &  $3.53 \pm 0.25$        \\
 $K_{\rm Ab}$ (\kms)                        &  $80.46 \pm 0.69$       \\
 $K_{\rm Bb}$ (\kms)                        &  $148.53 \pm 0.58$      \\ [1ex]
 $\phi_{\rm A,B}$ (degree)                  &  $76.12 \pm 0.24$       \\
 $\phi_{\rm A,AB}$ (degree)                 &  $92.34 \pm 0.19$       \\
 $\phi_{\rm B,AB}$ (degree)                 &  $59.68 \pm 0.20$       \\ [1ex]
 $(\mu_{\alpha}^*)_{\rm AB}$ (mas yr$^{-1}$)           &  $-175.88 \pm 0.46$     \\
 $(\mu_{\delta})_{\rm AB}$ (mas yr$^{-1}$)             &  $-99.28 \pm 0.43$ \\ [-1ex]
\enddata
\end{deluxetable}
\setlength{\tabcolsep}{6pt}

\begin{figure}
\epsscale{1.15}
\plotone{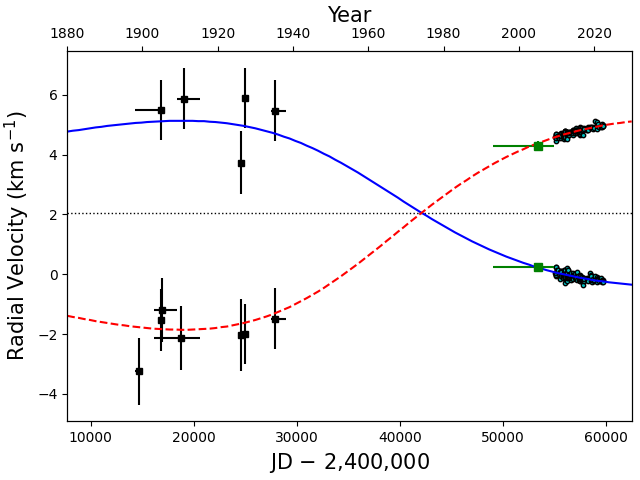}
\figcaption{Radial velocities in the Castor~AB orbit along with our
  model (solid blue line for the primary). Motion in the inner orbits
  has been subtracted. For TRES we represent the individual
  observations, whereas the DS measurements are shown by green
  squares at their median value and median time of observation. The
  corresponding velocity uncertainties are calculated from the median
  absolute deviations, and are approximately the same size as the
  symbols. The horizontal error bars for the DS represent the total time span of
  those observations. A dotted line marks the center-of-mass velocity
  of the quadruple system. As a check, we display also the
  historical velocities of both Castor~A and B from the sources
  described in the Appendix. In each case we subtract off the motion
  in the inner orbits, and plot the median value with its associated
  uncertainty, at the median epoch. The horizontal error bars
  represent the time span.\label{fig:outer_spec_orbit}}
\end{figure}

Plots of the radial velocities and CHARA observations in the inner
orbits (Figures~\ref{fig:charaA}--\ref{fig:RVinB}) and of the visual
observations in the outer orbit (Figure~\ref{fig:visual}) have been
shown previously. In Figure~\ref{fig:outer_spec_orbit} we now
represent our radial velocities in the outer orbit, after removal of
the motion of Castor~A and B in their respective inner spectroscopic
orbits. While the TRES observations are plotted individually, the less
precise DS measurements display significant scatter on the scale of
this figure, so to avoid clutter we represented them by a single point
marking the median value and its corresponding uncertainty. The much
more precise TRES velocities show clear evidence of the drift in the
center-of-mass velocities of both Castor~A and B.

As seen in the figure, our radial velocities cover only a small
section of the outer orbit (an interval of about 30~yr), but the velocity
semiamplitudes $K_{\rm A}$ and $K_{\rm B}$ are still well determined
because of all the other constraints used in our orbital analysis.
Nevertheless, it is of interest to compare our orbit with the
historical radial velocities from the literature going back more than
a century, which were not used in our fit. To do this we have taken
the most reliable sources as described in the Appendix, and subtracted
off the motion in the inner orbits according to our solution, leaving
only the component of the velocities in the outer orbit. The median
values for each of these historical sources are represented in
Figure~\ref{fig:outer_spec_orbit} as squares, and show remarkable
agreement with the model despite uncertainties in the velocity zero
points of those observations.

\section{Discussion}   
\label{sec:discussion}

\subsection{Comparison with stellar evolution models}

Our determination of the dynamical masses, angular diameters, and flux
ratios for Castor~Aa and Ba presents an opportunity to carry out a
comparison against models of stellar evolution, and to infer an age for
the system. The absolute radii follow directly from our measurement of
the uniform-disk angular diameters ($\phi_{\rm UD}$) from
Section~\ref{sec:interferometry}, a correction for limb darkening
based on the tabulations of \cite{Claret:2011}, and the distance. For the $H$-band
observations with MIRC-X, the linear limb-darkening coefficients used are
0.176 for Castor~Aa and 0.205 for Castor~Ba. For the $K$-band observations
with MYSTIC, the coefficients are 0.151 and 0.180, respectively. The
weighted average limb-darkened angular diameters are then $\phi_{\rm LD,Aa}
= 1.289 \pm 0.003$~mas and $\phi_{\rm LD,Ba} = 1.017 \pm 0.007$~mas. With our
distance to the system from Table~\ref{tab:results}, we obtain finally
$R_{\rm Aa} = 2.089 \pm 0.005~R_{\sun}$ and $R_{\rm Ba} = 1.648 \pm 0.011~R_{\sun}$.
Both the masses and the radii are formally determined to
better than 1\%.

\noindent

\begin{figure}
\epsscale{1.15}
\plotone{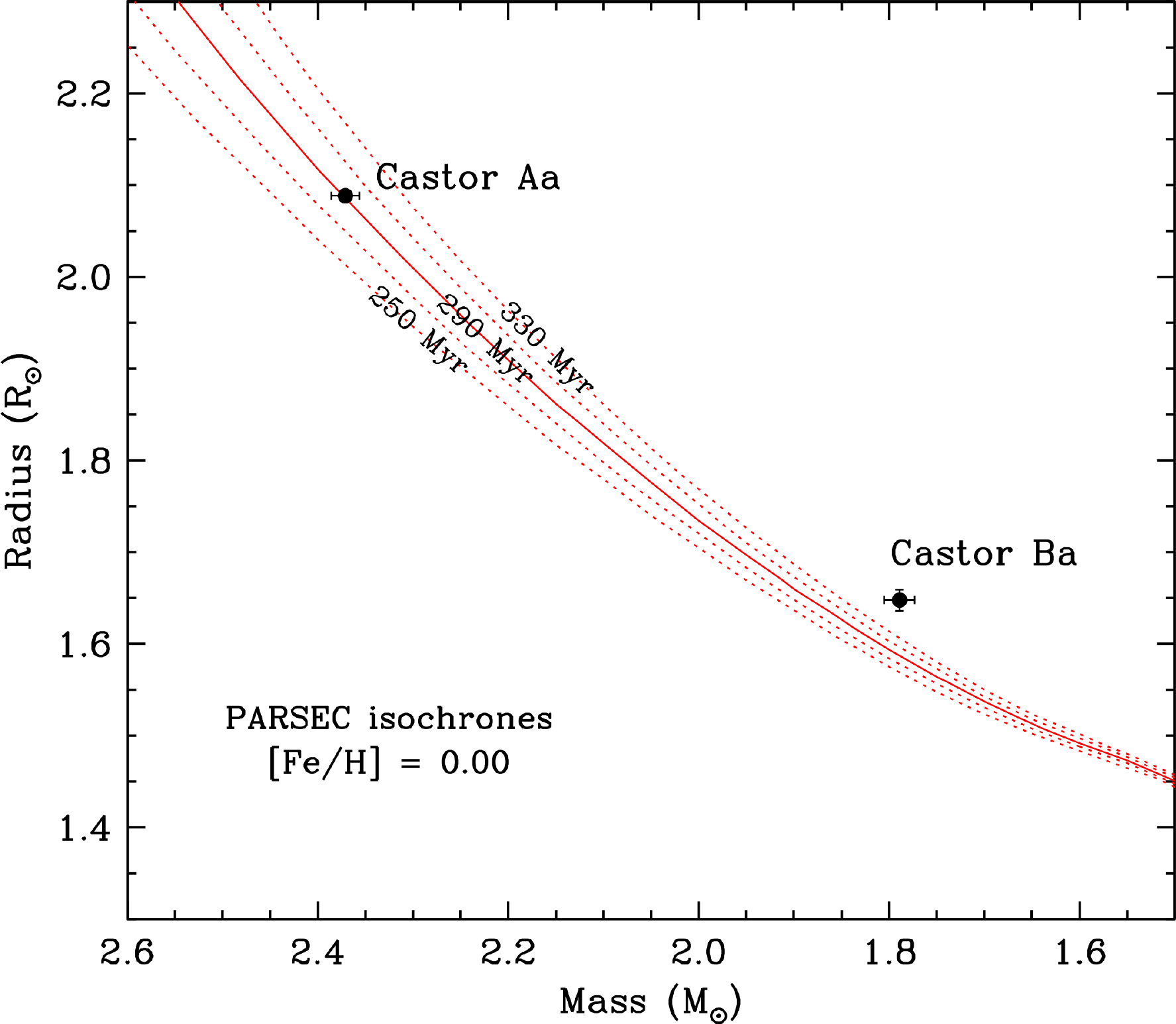}
\figcaption{Measured masses and radii of Castor~Aa and Ba shown
against solar-metallicity model isochrones from the PARSEC series of
\cite{Chen:2014}. Ages every 20~Myr are as labeled. The isochrone that
best fits the measurements for Castor~Aa, shown with a solid line, has
an age of 290~Myr.\label{fig:massradius}}
\end{figure}

The masses and radii are compared in Figure~\ref{fig:massradius} against
model isochrones from the PARSEC series of \cite{Chen:2014}. As
mentioned earlier, both objects are metallic-line A stars and they therefore
have anomalous surface abundances. For this comparison we used models with
solar metallicity that is more likely to be representative of their bulk
composition. We find that the models are not able to fit the properties of
both stars simultaneously at a single age. If we rely only on Castor~Aa,
the best fit age is 290~Myr and at this age the measured radius of Castor~Ba
appears about 5.5$\sigma$ ($\sim$4\%) larger than predicted for its mass. Considering the
observational challenges noted previously in gathering the CHARA measurements
for Castor~B, and their potential impact on the determination of the angular
diameter of its primary,
we have less confidence in the measured size for star Ba. Relying on it to
set the age would result in a far larger deviation for Castor~Aa in terms of its formal
uncertainty that seems implausible.

\begin{figure}
\epsscale{1.15}
\plotone{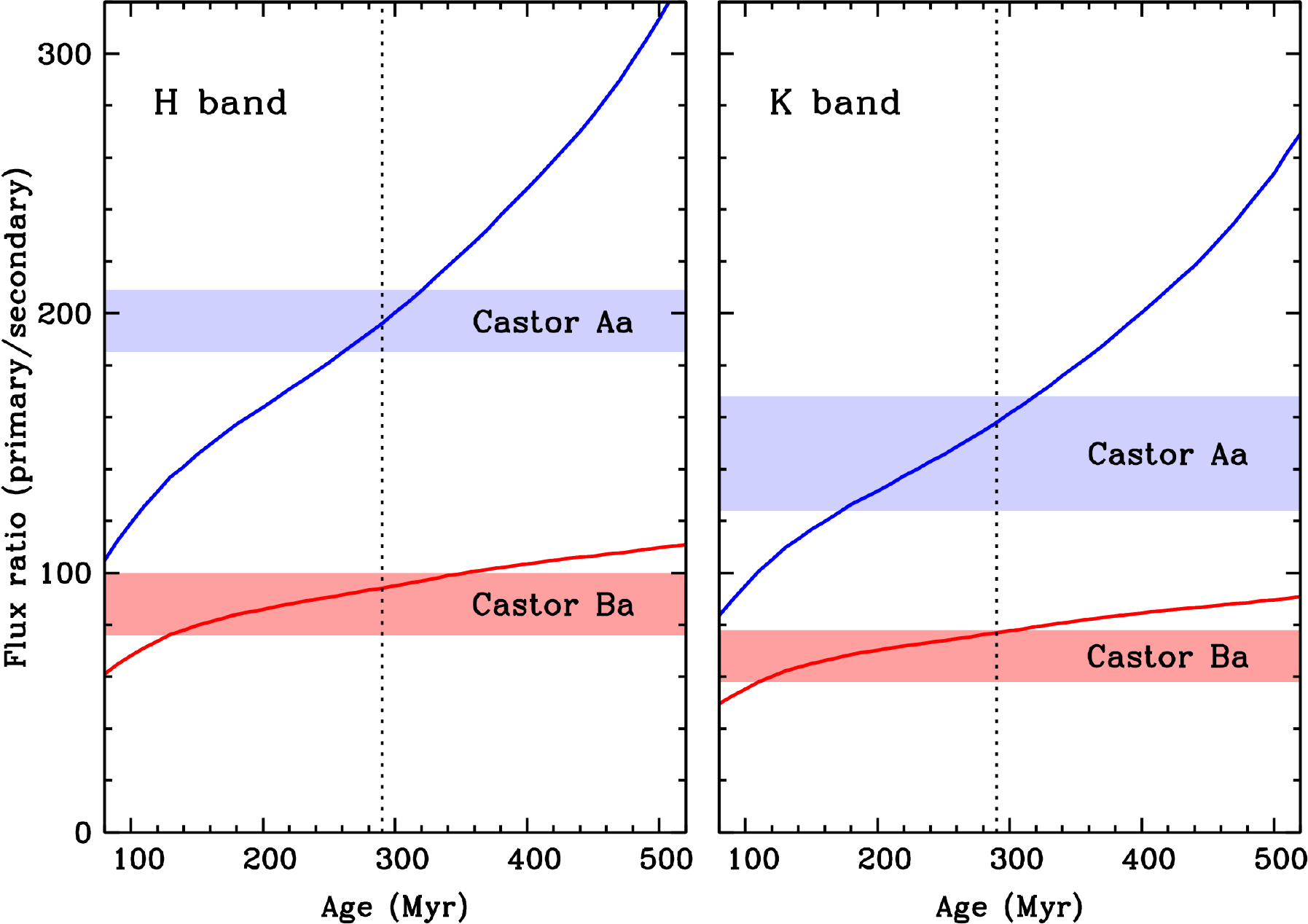}
\figcaption{Measured $H$-band and $K$-band primary to secondary flux
ratios for Castor~A and B with their uncertainties shown as shaded regions.
The solid lines represent the predicted run of the flux ratios as a function
of age in each system at the measured masses of the components,
according to the PARSEC models of \cite{Chen:2014}. Within the uncertainties
there is good agreement between the measurements and the expected flux ratios at the
age of 290~Myr (dotted line) that best fits the radius of Castor~Aa in
Figure~\ref{fig:massradius}.
\label{fig:fluxratio}}
\end{figure}

The measured $H$-band and $K$-band flux ratios for Castor~A and B, on the other hand, appear
consistent within their uncertainties with model predictions for the measured masses of the four stars
at the age of 290~Myr, as shown in Figure~\ref{fig:fluxratio}. They would
strongly disagree, particularly in the case of Castor~A, for the much older
age of 430~Myr needed to match the radius of Castor~Ba.

\subsection{The motion of Castor C (YY Gem)}
\label{sec:castorc}

Visual observers who recorded the relative position of Castor~A and B
occasionally also made measurements of the position angle and
separation of Castor~C with respect to either Castor~A or B. Similar
measurements were also made photographically. These observations,
referred to in the Washington Double Star Catalog as Castor~AC and BC,
respectively, contain useful information on the motion of Castor~C,
which is very slow on account of its wide separation from AB. We
estimate the orbital period to be roughly 14500~yr based on a total
mass of about 6.2~$M_{\sun}$ for the six-star system (see
Section~\ref{sec:analysis} for the mass of AB, and
\citealt{Kochukhov:2019} for YY~Gem).

Another independent indication of this slow motion can be obtained
from the difference in the proper motions of Castor~AB and C, although
this evidence is not particularly clear from the summary of
measurements given in Table~\ref{tab:pm}, which show considerable
scatter.  We have included in this listing the available
determinations for both objects that are most precise or that seem
more reliable to us (YY~Gem is not featured in the Hipparcos catalog
because it is too faint).  The Hipparcos (1997 and 2007) values for Castor~AB rest on
observations over an interval of only about three years, and as was
the case with the parallax results described earlier, they are
susceptible to errors caused by the orbital motion of Castor~A and B
around each other. The PPMX determinations are based on positions
spanning more than a century, but may still be affected.  Similarly
with the FK5 values. The fifth entry, formally very precise, is from
the recent US Naval Observatory Bright Star Astrometric Database
(UBAD), and is based on recent positional measurements combined with
the Hipparcos position from about 25~yr earlier.  The last entry for
Castor~AB is our updated p.m.\ from Hipparcos based on the orbital
analysis of Section~\ref{sec:analysis}, which is very different from
the original 1997 and 2007 results published by the mission.

\setlength{\tabcolsep}{1pt}  
\begin{deluxetable}{lccc}
\tablecaption{Proper Motion Determinations for Castor~AB and Castor~C\label{tab:pm}}
\tablehead{
\colhead{Source} &
\colhead{$\mu_{\alpha}^*$} &
\colhead{$\mu_{\delta}$} &
\colhead{Ref.}
\\
\colhead{} &
\colhead{(mas~yr$^{-1}$)} &
\colhead{(mas~yr$^{-1}$)} &
\colhead{}
}
\startdata
\multicolumn{4}{c}{Castor AB} \\
\noalign{\hrule} \\ [-1.5ex]
FK5 1988        &  $-171.56 \pm 0.37$\phn\phn\phs   &   $-98.70 \pm 0.41$\phn\phs      &  1 \\
Hipparcos 1997  &  $-206.33 \pm 1.60$\phn\phn\phs   &  $-148.18 \pm 1.47$\phn\phn\phs  &  2 \\
Hipparcos 2007  &  $-191.45 \pm 3.95$\phn\phn\phs   &  $-145.19 \pm 2.95$\phn\phn\phs  &  3 \\
PPMX 2008       &  $-169.69 \pm 0.40$\phn\phn\phs   &   $-98.29 \pm 0.40$\phn\phs      &  4 \\
UBAD 2022       &  $-193.77 \pm 0.12$\phn\phn\phs   & $-139.691 \pm 0.079$\phn\phn\phs &  5 \\
Hipparcos TD    &  $-175.88 \pm 0.46$\phn\phn\phs   &   $-99.28 \pm 0.43$\phn\phs      &  6 \\ [1ex]

\noalign{\hrule} \\ [-1.5ex]
\multicolumn{4}{c}{Castor C (YY Gem)} \\ [0.5ex]
\noalign{\hrule} \\ [-1.0ex]
PPMX 2008       &  $-200.25 \pm 1.20$\phn\phn\phs   &   $-92.52 \pm 1.30$\phn\phs      &  4 \\
UCAC4 2012      &   $-202.9 \pm 1.5$\phn\phn\phs    &    $-97.0 \pm 1.6$\phn\phs       &  7 \\
UCAC5 2017      &   $-204.5 \pm 3.1$\phn\phn\phs    &   $-100.9 \pm 3.1$\phn\phn\phs   &  8 \\
HSOY 2017       &  $-199.97 \pm 0.86$\phn\phn\phs   &   $-94.78 \pm 0.94$\phn\phs      &  9 \\
Gaia DR3        & $-201.406 \pm 0.029$\phn\phn\phs  &  $-97.000 \pm 0.025$\phn\phs     &  10 \\ [-1ex]
\enddata
\tablecomments{References in the last column: 
(1) \cite{Fricke:1988};
(2) \cite{ESA:1997}; 
(3) \cite{vanLeeuwen:2007}; 
(4) \cite{Roser:2008}; 
(5) \cite{Munn:2022};
(6) Re-analysis of the Hipparcos transit data (TD) from this paper (Table~\ref{tab:derived});
(7) \cite{Zacharias:2012};
(8) \cite{Zacharias:2017};
(9) \cite{Altmann:2017};
(10) \cite{Gaia:2022}, with uncertainties increased by factor of 1.37 following \cite{Brandt:2021}.
}
\end{deluxetable}
\setlength{\tabcolsep}{6pt}  

\begin{figure}
\epsscale{1.15}
\plotone{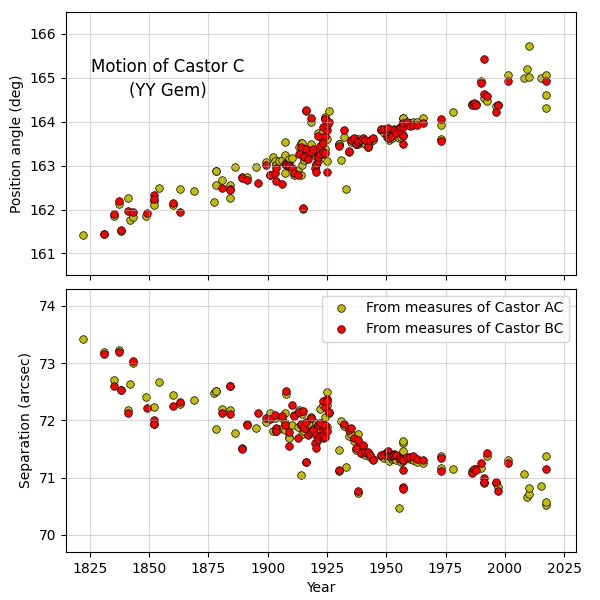}

\figcaption{Change in the position of Castor~C relative to the
  barycenter of Castor~AB from visual measures made over the past two
  centuries. The orbital motion of Castor~A and B with respect to their
  barycenter has been subtracted from the original measures of
  Castor~AC and BC, respectively. A few obvious outliers have been removed for
  clarity.\label{fig:yygem2}}

\end{figure}

\begin{figure}
\epsscale{1.15}
\plotone{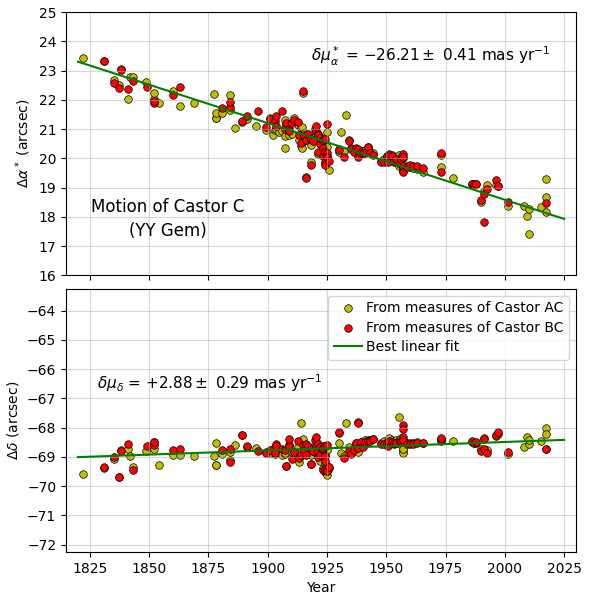}

\figcaption{Motion of Castor~C relative to the barycenter of Castor~AB
  in the R.A.\ and Dec.\ directions, from visual measures made over
  the past two centuries. The original measures of Castor~AC and BC
  have been reduced to the barycenter of AB.  Linear fits indicated by
  the green lines give the slopes $\delta\mu_{\alpha}^*$ and
  $\delta\mu_{\delta}$ shown in each panel, which are a direct measure
  of the difference in the proper motions of Castor~C and Castor~AB.
  A few outliers have been removed for clarity.\label{fig:yygem3}}

\end{figure}

Using the results from Section~\ref{sec:analysis}, we corrected the
more than 100 visual observations of Castor~AC for the motion of
component A around the center of mass of AB, and did the same for the
visual observations of Castor~BC. The adjusted values then reflect
only the motion of C relative to the barycenter of AB. They show
clearly that the position angles have increased by nearly
4\arcdeg\ over the past two centuries, and that the separations have
decreased by a little more than 2\arcsec\ (see
Figure~\ref{fig:yygem2}).  If we transform these measurements to
rectangular coordinates and assume the motion is linear to first
order, fits to the observations result in slopes in the R.A.\ and
Dec.\ directions of $-26.21 \pm 0.41$~mas~yr$^{-1}$ and $+2.88 \pm
0.29$~mas~yr$^{-1}$, respectively. These correspond to the true differences
$\delta\mu_{\alpha}^*$ and $\delta\mu_{\delta}$ between the
p.m.\ components for Castor~C and AB (Figure~\ref{fig:yygem3}), i.e.,
they reflect the orbital motion of C.

As a consistency check, we may use the high precision p.m.\ for
Castor~C from Gaia along with the above $\delta\mu_{\alpha}^*$ and
$\delta\mu_{\delta}$ values to infer the p.m.\ for Castor~AB. We
obtain $(\mu_{\alpha}^*)_{\rm AB} = -175.19 \pm 0.41$~mas~yr$^{-1}$ and
$(\mu_{\delta})_{\rm AB} = -99.88 \pm 0.30$~mas~yr$^{-1}$.  These
estimates are in fairly good agreement with our independent
determinations based on the Hipparcos transit data
(Table~\ref{tab:pm}).  The differences in each coordinate are at the
level of 1.1$\sigma$.

\subsection{Constraints on the Dynamics of the Castor system}
\label{sec:dynamics}

The sextuple system of Castor has been the subject of several
dynamical studies over the past three decades. Noting that the typical
separation of the three spectroscopic subsystems from each other is
large, \cite{Anosova:1989} treated Castor as a triple system (A, B, C)
for dynamical purposes, and showed based on the limited information
then available that it is gravitationally bound. Subsequent studies,
still handicapped by the limited knowledge of many of the orbital
properties, have investigated the stability of the various subsystems,
and the long-term evolution of some of their orbital elements
\citep{Beust:2003, Andrade:2015, Matvienko:2015, Docobo:2016}. The
latter study focused on the possibility that some of the subsystems
are undergoing Kozai-Lidov (KL) cycles \citep{Kozai:1962, Lidov:1962},
a mechanism that transfers angular momentum between the inner and
outer orbits causing oscillations in the inner eccentricities and
inclination angles. These works have concluded that Castor~A, B, and
AB are dynamically stable, and that Castor~A is currently undergoing
KL cycles, and likely experiencing apsidal motion.

The present study represents a major step forward in our knowledge of
the system. The masses of all six stars are now individually
determined to better than 1\%, and their sum to better than 0.5\%
(smaller because of correlations among them). We have now also
established the full 3D orbits of Castor~A, Castor~B, and Castor~AB,
with all orientation angles ($i$, $\omega$, $\Omega$) being measured
to better than 0\fdg5. This means that the mutual inclination angles
of their orbital planes can also be determined to high precision, and
are known to better than 0\fdg5 as well (Table~\ref{tab:derived}).
The orbit of Castor~A is nearly at right angles to that of AB ($\sim$92\arcdeg,
formally in retrograde motion), while the orbit of B is inclined
about 60\arcdeg\ to that of AB. 
Dynamical stability criteria that account for the mutual inclination
angles, such as that of \cite{Mardling:2001} as formulated by
\cite{He:2018}, or that of \cite{Myllari:2018}, indicate that
neither Castor~A nor Castor~B are in danger of being disrupted,
as found also by earlier studies. Here we have assumed, for the purpose
of applying these criteria developed for triple systems,
that each binary feels the other as if it were a point mass.
Regarding the retrograde motion of Castor~A, both criteria suggest
that retrograde orbits are actually more stable than prograde
orbits for a given ratio of pericenter distance of the outer orbit
to the semimajor axis of the inner orbit.

While it is beyond the scope of this paper, it would clearly be
worthwhile to revisit some of the more sophisticated dynamical studies
of Castor mentioned above, armed
with the much more complete information now available. The main
limitation, however, remains our very poor knowledge of the orbit of
Castor~C. Our discussion in the previous section and the precise
determination of the relative motion between C and AB are an
improvement over the analysis of \cite{Matvienko:2015}, but the
measured arc of that orbit is still very small, and will remain so for
centuries. Nevertheless, the fact that all masses are well known
provides a useful and previously unavailable constraint on the scale
of that wide orbit, in the form of the ratio $a^{\prime\prime\,3}_{\rm
  AB,C}/P_{\rm AB,C}^2$, which is now established with a precision
better than 0.5\%.

An additional limitation has to do with the radial velocities. Having
precise and accurate values for the center-of-mass velocities of
Castor~AB and Castor~C would be of considerable help in constraining
their relative motion.  Unfortunately, it is difficult to assess the
accuracy of our determination of $\gamma$ for Castor~AB in
Table~\ref{tab:results} because of the potential for systematic errors
caused by the chemical anomalies mentioned earlier.
Based on the range of values for the
offsets $\Delta$ in the first block of Table~\ref{tab:results}, we do
not believe our result of +2.057~\kms\ should deviate from the true
value of $\gamma_{\rm AB}$ by more than 1~\kms. The historical radial
velocities for Castor described in the Appendix provide some support
for this statement.  As discussed in detail there, the $\gamma_{\rm
  AB}$ estimates one may derive from those observations, transformed
as closely as possible to the same velocity system as ours (see the
Appendix), are surprisingly consistent and range between +2.36
and +2.42~\kms, or roughly 0.4~\kms\ higher than ours. However,
additional slight differences in the velocity zero points of those
determinations compared to ours cannot be ruled out, and are always of
concern when discussing velocity differences of order 1~\kms\ or less.

There is also some scatter in the available velocities of the center
of mass of Castor~C. \cite{Torres:2002} reported $\gamma_{\rm C} =
+0.68 \pm 0.26$~\kms\ (adjusted here to include a correction of
+0.14~\kms\ to the IAU system as described in
Section~\ref{sec:spectroscopy}), while \cite{Segransan:2000} estimated
$\gamma_{\rm C} = +1.97 \pm 0.24$~\kms. The more recent determination
by \cite{Kochukhov:2019} gave a more precise value of $\gamma_{\rm C}
= +2.287 \pm 0.038$~\kms. The first of these determinations should be
on a similar velocity system as ours, although it too may be affected
by template mismatch to some extent because the synthetic templates
used may not provide a perfect representation of stars as cool as
those in Castor~C, with nearly identical components of spectral type
\ion{M1}{5}. On the other hand, the velocity zero points for the other
two determinations are not described in those studies, but are likely
to be slightly different from ours.  Given that our velocities have
the gravitational redshift and convective blueshift of the Sun
subtracted out (Section~\ref{sec:spectroscopy}), and theirs presumably
do not, at the very least the velocities from \cite{Segransan:2000}
and \cite{Kochukhov:2019} probably require a slight offset of
$-0.29~\kms$ to place them on the same standing as ours. See the
Appendix for a discussion of this issue. If we adopt the most precise
of those estimates at face value and reduce it to the system of our
velocities (giving $\gamma_{\rm C} = +2.00$), we conclude that the
difference between the centers of mass of Castor AB and C is probably
on the order of just 0.5~\kms, if not less. At this level other
physical effects come into play that are not necessarily negligible.
For example, because the spectral types of Castor~AB and C are so
different, an additional contribution of $\sim$0.1~\kms\ stems from
the difference in the respective gravitational redshifts (that of
Castor~C being lower). A further contribution that is much more
difficult to quantify comes from the difference in the convective
blueshifts, which are poorly known for both A stars and M stars.

\section{Conclusions}
\label{sec:conclusions}

Using the CHARA array, we have resolved the spectroscopic companions of
both Castor~A and B for the first time, thanks to the excellent capabilities
of the MIRC-X and MYSTIC beam combiners. The challenge in this case was not so much the
small angular separations ($\sim$1~mas and larger), but rather the large
contrast ratios between the A-type primaries and the M-type secondaries:
star Ab contributes a mere 0.5\% of the $H$-band flux of Aa, and star Bb only about
1\% of the light of Ba. The contributions are only slightly larger in the $K$ band.
Both of these close companions are identical stars of spectral type M2 or M3.

While Castor~A and B are too bright for Gaia to have determined a trigonometric
parallax for the system (as of DR3), the presence of the fainter and wide physical
companion Castor~C allowed that valuable piece of information to be obtained
with high precision. With this additional constraint, and our extensive
radial velocity monitoring over a nearly 30 yr time span, we have established
the masses of all four stars in Castor~A and B to better than 1\%. Measurements
of the relative positions of A and B gathered by visual observers since the early 1700's, along with
our velocities, have then enabled us to determine the 3D orbits for Castor~A, B,
and AB. They are nowhere near coplanar, which is not particularly unusual for
systems with wide outer orbits \citep[see, e.g.,][]{Tokovinin:2017}, but
they are dynamically stable.

The masses of all six stars in this remarkable nearby system are now well known,
along with all of their orbital properties except for those of $\sim$14,000~yr
path of Castor~C around the AB
quadruple. We have provided additional constraints on that orbit that should
help future studies of the stability and evolution of the ensemble. Based
on the measured angular diameter of the primary of Castor~A, we infer
an age for the system from current stellar evolution models of 290~Myr.

\appendix
\label{sec:appendix}
\section{Historical Radial Velocity Measurements}

Here we summarize the historical radial velocity measurements of
Castor~A and B of which we are aware, which began more than a century
ago. Many of these observations are remarkably good for the time, and
fully support the outer spectroscopic orbit described in the main
text.  We describe the sources of radial velocity measurements for each
component separately, beginning with Castor~B, the fainter star of the
visual pair, whose binary nature was discovered first.

\subsection{Castor B}

The announcement that Castor B is a 2.9 day spectroscopic binary was
made by \cite{Belopolsky:1897}, in a paper reporting 32 photographic
RV measurements obtained in 1896 at the Pulkovo Observatory. The dates
(to two decimal places in days) were given in Pulkovo mean time (two
hours later than GMT), and the velocity measurements were made
originally in units of German geographical miles, equivalent to
7.42~km. They were converted for publication to \kms\ by the editors of
the journal. The following year the same author \citep{Belopolsky:1898}
published 14 additional measurements obtained in 1894 (2) and 1897 (12),
which were reported in geographical miles and the same
time units as before, but to only one decimal place. Then two years
later \cite{Belopolsky:1900} republished 21 of the 1896 measurements,
along with 18 from 1898 and 21 from 1899 in a study addressing the
possibility of apsidal motion in Castor~B (which has since been
dismissed). Two decimal places were given for the dates, and a few of
the velocities and dates are slightly different from those listed in
other tables of the same paper, for reasons that are unclear. B\'elopolsky's measurements were
reproduced by \cite{Curtis:1906} in units of \kms, with the times
of observation converted to Julian dates (given to two decimal
places). Once again some of the velocities are different from those in
the earlier papers by B\'elopolsky.  \cite{Curtis:1906} appears to
have had access to B\'elopolsky's original measurements, even some
unpublished ones, as on several of the 1899 nights he listed more
than one measurement where B\'elopolsky's had simply published an
average. Of the 118 individual B\'elopolsky velocity measurements
(1894, 1896--1899) as reported by \cite{Curtis:1906}, the ones prior
to 1898 appear to have more scatter and are systematically offset,
perhaps due to changes in the spectrograph around 1897. We have
chosen not to use them for our purposes.

Two observations from 1896 reported by \cite{Newall:1897}
were described as representing velocity differences between Castor~B
and Castor~A. They were made at the Royal Greenwich Observatory on
consecutive nights, without a comparison spectrum, as indicated by the
author.  \cite{Belopolsky:1898} expressed difficulty in reconciling
those measurements with his orbit, assuming Castor~A and Castor~B have
the same systemic velocity and that the velocity of Castor~A is
constant. We have not been able to make sense of them either, even
after accounting for the motion in the updated inner orbits of both
Castor~A and Castor~B, as well as the outer orbit. We have therefore
disregarded these measurements.

Some years later \cite{Lehmann:1924} published 147 additional velocity
measurements of Castor~B from photographic plates taken by
B\'elopolsky between 1903 and 1917. The dates were given to three
decimal places, in Pulkovo mean time. Detailed notes about the quality
of the plates were provided, and we used them to remove a few
observations of poor quality.  As the author noted, the measurements
from 1916 and 1917 all appear to be significantly offset toward
negative values relative to the earlier ones, by an amount that we
estimate to be 7--8~\kms\ based on our own analysis. The reasons for
this are unclear. We note that while the offset happens to be very
near one German geographical mile (7.42~\kms, i.e., one unit of
measurement as used originally by B\'elopolsky), it is more likely
related to changes in the instrument noted by \cite{Lehmann:1924}. We
have retained only the 1903--1915 measurements here.

Radial velocities from 32 spectroscopic observations of Castor~B
obtained in 1904--1905 at the Lick Observatory were reported in the
same paper by \cite{Curtis:1906} mentioned earlier. They are of
excellent quality, and we adopted them as published except for a minor
velocity zero point adjustment of $-0.2$~\kms, following
\cite{Campbell:1928}. Additional RV measurements from the Dominion
Astrophysical Observatory made between 1926 and 1927 were published by
\cite{Barlow:1928}. However, the velocity zero point of these 42 measurements
is uncertain, as the author stated the plate measurements
were made relative to another taken as the standard, but did not provide any further
information about that standard. Consequently, we have
elected not to use them. Two additional RV measurements from 1926 made
on the same night at the same observatory were published separately by
\cite{Harper:1937}, and appear to be of better quality. We have made
use of them here, along with a single RV measurement in 1927 from the
Yerkes Observatory that was reported by \cite{Frost:1929}.

Another important series of RV measurements from the Lick Observatory
was published by \cite{Vinter-Hansen:1940}. These observations were
obtained between 1934 and 1938, and are also of very good quality
judging by the rms residuals from an orbital solution. 

Very few other velocity measurements of Castor~B have appeared in the
literature since. One short series of five observations from the
Greenwich Observatory (five RVs from 1961--1962) was published by
\cite{Palmer:1968}, but unfortunately the measurements are too poor to
be of use, as they have internal errors of 4--6~\kms. Two additional velocities
from a single night at the Kitt Peak National Observatory in 1970 by
\cite{Abt:1980} are also too poor to be helpful.  To our knowledge
there are no published velocities from the Mount Wilson Observatory. A
single RV from 2007 reported by \cite{Auriere:2010} is also not
useful, as the authors caution the zero point is very uncertain. It
also overlaps in time with ours, so it would not contribute
significantly.

\subsection{Castor A}

The discovery that Castor~A is a spectroscopic binary with a 9.2 day
period was made in 1904 by H.\ D.\ Curtis at the Lick Observatory, and
first mentioned in print by \cite{Campbell:1905a} and
\cite{Campbell:1905b}.  The latter paper listed 25 preliminary RV
measurements (1897--1905), which were revised and augmented to 49 the
next near by \cite{Curtis:1906}. While in some cases multiple
observations on the same night were also presented as averages, we
have used only the original 49 measurements here. As in the case of
Castor~B, we have adjusted these values by $-0.2$~\kms\ following
\cite{Campbell:1928}.

A series of photographic plates of Castor~A secured by B\'elopolsky at
the Pulkovo Observatory between 1909 and 1916 were measured and
published by \cite{Rossowskaya:1924}, who reported 38 RV
determinations. The dates are expressed in Pulkovo mean time, and the
velocities (reported in \kms, to two decimal places) are of excellent
quality. A publication by \cite{Barlow:1928} gave 48 additional
velocity measurements from 1926--1927 made at the Dominion
Astrophysical Observatory, but as mentioned earlier for Castor~B, the
uncertain zero point makes these values of little use for our
analysis.

Other RV measurements for Castor~A include two by \cite{Harper:1937}
from 1926 (Dominion Astrophysical Observatory), one by
\cite{Frost:1929} from 1927 (Yerkes Observatory), and seven made at
the Greenwich Observatory from 1960--1962 by \cite{Palmer:1968}. The
latter are very poor, as was the case for the measurements of Castor~B
from the same source, and we have not used them. Similarly with the
single 1970 velocity measurement by \citep{Abt:1980}. The catalog of
individual radial velocity measurements from the Mount Wilson
Observatory \citep{Abt:1970} contains no entries for Castor~A.

The only other data set for Castor~A we are aware of is that of
\cite{Vinter-Hansen:1940}, who published 48 good-quality velocities
from the Lick Observatory made between 1934 and 1938. Together with
the observations of Castor~B made by the same authors, these two are
the most recent extensive series of RVs available for either component
of Castor.

\subsection{Using the Historical Velocities}

Most of the main sources of radial velocities for Castor described
above appear to have more or less consistent zero points, although
small discrepancies are bound to be present due to the peculiarities
of each instrument and the different ways in which the measurements
were made. It is not possible to place the measurements from each of
these historical sources accurately on the same absolute frame of
reference as our own velocities. For this reason, we have opted
against incorporating them into our orbital solution to constrain the
outer orbit. Instead, we used them as a check on the velocity
semiamplitudes $K_{\rm A}$ and $K_{\rm B}$ and the center of mass
velocity $\gamma_{\rm AB}$ that we derived from our global solution.
This comparison is shown in Figure~\ref{fig:outer_spec_orbit} in the main
text.

By coincidence, all of the historical RV measurements happen to be
from a time when the velocities of Castor~A and B in the outer orbit
were near their extremes, and were therefore not changing much. In
order to provide a representative value of the center-of-mass velocity
for each visual component in the AB pair from these sources, we first
subtracted the motion in the inner binaries from the individual velocities
using the parameters from our global solution in Table~\ref{tab:results}).
We then calculated the median of the resulting residuals for each data
set, along with its formal uncertainty. We used the median rather than
the mean as a more robust estimate against outliers.  The results are
given in Table~\ref{tab:historicalA} and Table~\ref{tab:historicalB},
where we list also the interval in years, the median Julian date, and
the number of velocity measurements from each source.

We have arbitrarily increased the formal errors listed in these tables
by adding 1~\kms\ in quadrature to account for possible differences in
the instrumental velocity zero points from the different sources.
These larger errors are the ones shown in
Figure~\ref{fig:outer_spec_orbit}. Additionally, we have applied a
small but systematic shift to all historical velocities in the amount
of $-0.29$~\kms\ (not included in the above tables). This is to account for
the fact that the literature values are affected by the gravitational
redshift and any convective blueshift in Castor~A and B, whereas by
construction our velocities (and the spectroscopic orbit derived from them) are only affected by the difference
between those two effects in Castor and in the Sun. To be explicit,
the measured historical velocities may be expressed as $RV_{\rm
  obs}{\rm (hist)} = RV_{\rm true} + GR + CB$, where $GR$ and $CB$ are
the gravitational redshift and convective blueshift, respectively.
The CfA velocities, on the other hand, can be written as $RV_{\rm obs}{\rm (CfA)} = RV_{\rm
  true} + (GR - GR_{\sun}) + (CB - CB_{\sun})$. The difference is then
$RV_{\rm obs}{\rm (hist)} - RV_{\rm obs}{\rm (CfA)} = GR_{\sun} +
CB_{\sun} = 0.63 - 0.34 = +0.29~\kms$. The value of $CB_{\sun} =
-0.34~\kms$ was taken from Figure~3 of \cite{Meunier:2017}.
The agreement in Figure~\ref{fig:outer_spec_orbit} between the
historical RVs and the predicted velocity curves from our global
solution is surprisingly good, considering that some of those
measurements were made more than a century ago.

\setlength{\tabcolsep}{6pt}
\begin{deluxetable}{llcccc}
\tablecaption{Center-of-Mass Velocities for Castor~A from Historical Measurements\label{tab:historicalA}}
\tablehead{
\colhead{Source} &
\colhead{Observatory} &
\colhead{Interval} &
\colhead{Median JD} &
\colhead{$N_{\rm RV}$} &
\colhead{Median $RV_{\rm A}$}
\\
\colhead{} &
\colhead{} &
\colhead{} &
\colhead{(2,400,000+)} &
\colhead{} &
\colhead{(\kms)}
}
\startdata
\cite{Curtis:1906}         & Lick                    & 1897--1905  &  16857  &   49  &  $+5.79 \pm 0.15$ \\
\cite{Rossowskaya:1924}\tablenotemark{a} & Pulkovo   & 1909--1915  &  19074  &   38  &  $+6.17 \pm 0.25$ \\
\cite{Harper:1937}         & Dominion                & 1926        &  24561  &    2  &  $+4.02 \pm 0.31$ \\
\cite{Frost:1929}          & Yerkes                  & 1927        &  24936  &    1  &  $+6.18$ \\
\cite{Vinter-Hansen:1940}  & Lick                    & 1934--1938  &  27849  &   48  &  $+5.77 \pm 0.12$ \\ [-1ex]
\enddata
\tablecomments{A systematic shift of $-0.29~\kms$, not included here, should be applied to
  all of these velocities for the comparison with the curves in
  Figure~\ref{fig:outer_spec_orbit} (see the text).  We also
  conservatively increase the internal errors listed here by adding
  1~\kms\ in quadrature, to account for additional zero point
  differences.}
\tablenotetext{a}{Measurements from 1916--1917 omitted (see the text).}
\end{deluxetable}
\setlength{\tabcolsep}{6pt}  

\setlength{\tabcolsep}{6pt}
\begin{deluxetable}{llcccc}
\tablecaption{Center-of-Mass Velocities for Castor~B from Historical Measurements\label{tab:historicalB}}
\tablehead{
\colhead{Source} &
\colhead{Observatory} &
\colhead{Interval} &
\colhead{Median JD} &
\colhead{$N_{\rm RV}$} &
\colhead{Median $RV_{\rm B}$}
\\
\colhead{} &
\colhead{} &
\colhead{} &
\colhead{(2,400,000+)} &
\colhead{} &
\colhead{(\kms)}
}
\startdata
\cite{Curtis:1906}\tablenotemark{a}  & Pulkovo   & 1898--1899  &  14687  &   72  &  $-2.96 \pm 0.48$ \\
\cite{Curtis:1906}         & Lick                & 1904--1905  &  16850  &   32  &  $-1.26 \pm 0.26$ \\
\cite{Lehmann:1924}        & Pulkovo             & 1903--1915  &  18752  &  115  &  $-1.83 \pm 0.39$ \\
\cite{Harper:1937}         & Dominion            & 1926        &  24561  &    2  &  $-1.74 \pm 0.66$ \\
\cite{Frost:1929}          & Yerkes              & 1927        &  24936  &    1  &  $-1.70$ \\
\cite{Vinter-Hansen:1940}  & Lick                & 1934--1938  &  27930  &   44  &  $-1.19 \pm 0.19$ \\ [-1ex]
\enddata
\tablecomments{A systematic shift of $-0.29~\kms$, not included here, should be applied to
  all of these velocities for the comparison with the curves in
  Figure~\ref{fig:outer_spec_orbit} (see the text). We also
  conservatively increase the internal errors listed here by adding
  1~\kms\ in quadrature, to account for additional zero point
  differences.}
\tablenotetext{a}{Measurements made originally by
  \cite{Belopolsky:1897}, \cite{Belopolsky:1898}, and
  \cite{Belopolsky:1900}, with additional unpublished RVs by the same
  author. RVs prior to 1898 have a larger scatter and have been
  omitted.}
\end{deluxetable}
\setlength{\tabcolsep}{6pt}  

The historical velocities can be used in a different way to provide
independent estimates of the center-of-mass velocity of the quadruple system Castor~AB.
This quantity is of relevance for constraining the orbit of Castor~C
(Section~\ref{sec:dynamics}), which is currently very poorly defined
because of its very long orbital period of $\sim$14500~yr. For each of
the data sets in Tables~\ref{tab:historicalA} and
\ref{tab:historicalB} from the same observatory and at similar epochs we
combined the center-of-mass velocities of Castor~A and B with the
expression
\begin{equation}
\gamma_{\rm AB} = \frac{\gamma_{\rm A}M_{\rm A}+\gamma_{\rm B}M_{\rm B}}
{M_{\rm A} + M_{\rm B}}~,
\end{equation}
where $M_{\rm A}$ and $M_{\rm B}$ are the masses from
Table~\ref{tab:derived}. The results, collected in
Table~\ref{tab:gamma}, show remarkable agreement considering their diverse nature except for the
estimate from \cite{Harper:1937}, which is lower. We note, however,
that this author explicitly mentioned a systematic difference between
the velocities from the Dominion Astrophysical Observatory and those
from the Lick Observatory between $-1.0$ and $-1.5$~\kms, with the ones from
Dominion being lower. If an offset in the middle of that range were
applied to the result from Table~\ref{tab:gamma}, it becomes
+2.44~\kms, in excellent agreement with the others.

\setlength{\tabcolsep}{6pt}
\begin{deluxetable}{lcccc}
\tablecaption{Center-of-Mass Velocities for Castor~AB from Historical Measurements\label{tab:gamma}}
\tablehead{
\colhead{Observatory} &
\colhead{Mean Epochs (A and B)} &
\colhead{$N_{\rm A}$} &
\colhead{$N_{\rm B}$} &
\colhead{$\gamma_{\rm AB}$ (~\kms)}
}
\startdata
Lick     &  1905.08 / 1905.06  &  49  &  32  &  +2.39 \\
Pulkovo  &  1911.15 / 1910.27  &  38  & 115  &  +2.36 \\
Dominion &  1926.18 / 1926.18  &   2  &   2  &  \phm{\tablenotemark{a}}+1.19\tablenotemark{a} \\
Yerkes   &  1927.20 / 1927.20  &   1  &   1  &  +2.42 \\
Lick     &  1935.18 / 1935.40  &  48  &  44  &  +2.41 \\ [-1ex]
\enddata
\tablecomments{In calculating these center-of-mass velocities, a systematic shift of $-0.29~\kms$ has been applied to
  the velocities of Castor~A and B from Tables~\ref{tab:historicalA}
  and \ref{tab:historicalB} to place them on the same system as the
  new velocities in this paper as closely as possible (see the text).}
\tablenotetext{a}{This discrepant value is explained by the systematic
  difference that exists between the Dominion velocities and those
  from Lick, as described by \cite{Harper:1937}. An approximate correction
  for that difference following that author would change it to
  +2.44~\kms, bringing it in line with the others (see the text).}
\end{deluxetable}
\setlength{\tabcolsep}{6pt}  


\begin{acknowledgements}

This work is based upon observations obtained with the Georgia State University Center for High Angular Resolution Astronomy Array at Mount Wilson Observatory.  The CHARA Array is supported by the National Science Foundation under Grant No.\ AST-1636624 and AST-2034336.  Institutional support has been provided from the GSU College of Arts and Sciences and the GSU Office of the Vice President for Research and Economic Development. MIRC-X received funding from the European Research Council (ERC) under the European Union's Horizon 2020 research and innovation programme (Grant No.\ 639889). JDM acknowledges funding for the development of MIRC-X (NASA-XRP NNX16AD43G, NSF-AST 1909165) and MYSTIC (NSF-ATI 1506540, NSF-AST 1909165). Time at the CHARA Array was granted through the NOIRLab community access program (NOIRLab PropID: 2021A-0008, 2021B-0009; PI: G.\ Torres). This research has made use of the Jean-Marie Mariotti Center Aspro and SearchCal services.
SK and CLD  acknowledge support by the European Research Council (ERC Starting Grant, No.\ 639889 and ERC Consolidator Grant, No.\ 101003096), and STFC Consolidated Grant (ST/V000721/1). AL received funding from STFC studentship No.\ 630008203.

The spectroscopic observations of Castor at the CfA were gathered with
the expert help of P.\ Berlind, M.\ Calkins, J.\ Caruso, G.\ Esquerdo,
D.\ Latham, R.\ Stefanik, and J.\ Zajac. We thank R.\ Davis and
J.\ Mink for maintaining the CfA echelle databases. We are also
grateful to M.\ McEachern (Wolbach Library) for her valuable
assistance in locating and providing copies of some of the historical
papers for Castor, and to M.\ Badenas Agusti for her help in the early
stages of this project. The anonymous referee is thanked as well for
helpful comments.

The research has made use of the SIMBAD and VizieR databases, operated
at the CDS, Strasbourg, France, of NASA's Astrophysics Data System
Abstract Service, and of the Washington Double Star Catalog maintained
at the U.S.\ Naval Observatory.
The work has also made use of data from the European Space Agency
(ESA) mission Gaia (\url{https://www.cosmos.esa.int/gaia}), processed
by the Gaia Data Processing and Analysis Consortium (DPAC,
\url{https://www.cosmos.esa.int/web/gaia/dpac/consortium}). Funding
for the DPAC has been provided by national institutions, in particular
the institutions participating in the Gaia Multilateral Agreement. The
computational resources used for this research include the Smithsonian
High Performance Cluster (SI/HPC), Smithsonian Institution
(\url{https://doi.org/10.25572/SIHPC}).
\end{acknowledgements}

\end{document}